\documentclass[nonlinenumber]{aa}  

\usepackage{algorithm}
\usepackage{algorithmic}
\usepackage{caption}
\usepackage{booktabs} 
\usepackage{amsmath}
\usepackage{makecell}
\usepackage{tikz}
\usepackage{hyperref}
\usepackage{lineno}

\usepackage{graphicx}
\usepackage{txfonts}
\usepackage{lipsum}
\usepackage{subcaption}         
\usepackage{lscape}             
\usepackage{placeins}           
                                
\begin{document}

   \title{DELOS: Contrastive Deep Learning for Low-SNR Blind Transit Searches in Kepler Photometry}

%
%
%

    \author{Qingtian Liu\inst{1,2}\email{qtxplorer@shao.ac.cn}
    \and Jian Ge\inst{1}\corrauth{jge@shao.ac.cn} 
    \and XingChen Yan\inst{1,2}\email{yanxingchen0@gmail.com}
    \and Kevin Willis\inst{3}\email{kevin.w.willis@gmail.com}
    \and Xinyu Yao\inst{1}\email{1034432842@qq.com}
    \and QuanQuan Hu\inst{1,2}\email{huqq1917@outlook.com}
    \and Jiapeng Zhu\inst{1}\email{jpzhu@smail.nju.edu.cn}
    }

   \institute{Shanghai Astronomical Observatory, Chinese Academy of Sciences, 200030, Shanghai, China
   \and School of Astronomy and Space Sciences, University of Chinese Academy of Sciences, 101408, Beijing, China
   \and Science Talent Training Center, Gainesville, 32606, FL, USA}

   \date{Received September 30, 20XX}


\abstract
  {Detecting shallow intermediate-to-long-period transits in \textit{Kepler} photometry remains challenging because low-SNR signals are easily missed by conventional period-search methods.}
  {We present \textbf{DE}tection in phase-folded \textbf{L}ight curves with c\textbf{O}ntrastive \textbf{S}coring (\textbf{DELOS}), a deep-learning framework that uses contrastive scoring to perform blind searches for shallow transits in \textit{Kepler} photometry.}
  {DELOS combines GPU-accelerated phase folding, optimized phase binning, and a custom one-dimensional convolutional encoder to assign a transit-likeness score to each folded light curve, thereby producing a score periodogram over trial periods without relying on pre-detected threshold-crossing events. 
Focusing on intermediate-to-long-period signals with orbital periods of 100–150 days, DELOS was trained on 20 million synthetic light curves generated with realistic transit models and Kepler-like noise properties, achieving a validation accuracy of 99.3\% on the synthetic validation set.}
  {In controlled injection-recovery experiments, DELOS improves the combined precision–recall performance by 15.5\% relative to Box-fitting Least Squares (BLS) and 11.25\% relative to Transit Least Squares (TLS) in the low Signal-to-Noise Ratios (low-SNR) regime. It also accelerates the search by factors of approximately 3–5 and 74–80 compared with BLS and TLS, respectively.
Applied to a selected \textit{Kepler} validation sample, DELOS recovered all known shallow intermediate-to-long-period transit signals in the tested period range.}
  {These results demonstrate that DELOS provides an efficient and sensitive framework for low-SNR transit searches and represents a practical step toward future searches for longer-period terrestrial planets in \textit{Kepler}, \textit{K2}, \textit{TESS}, \textit{PLATO}, and \textit{Earth 2.0} data. Accordingly, this work is intended as a methodological development and validation study, with the detailed astrophysical validation of newly identified candidates deferred to future work.}


   \maketitle


\section{Introduction}\label{sec:1}
Since the discovery of the first exoplanet orbiting a solar-type star \citep{Mayor1995}, over 6,000 exoplanets have been discovered, with several thousand additional candidates yet to be confirmed.
The primary methods for detecting these exoplanets include transiting \citep[e.g.,][]{Charbonneau_2000}, radial velocity \citep[e.g.,][]{1988ApJ...331..902C}, direct imaging \citep[e.g.,][]{refId0}, and microlensing \citep[e.g.,][]{Beaulieu2006}.
Among these methods, transit photometry has been the most productive, contributing to the discovery of over 4,600 exoplanets according to \href{https://exoplanetarchive.ipac.caltech.edu/docs/counts_detail.html}{NASA Exoplanet Archive statistics}.

Initially, potential transit signals were detected by ground-based photometric observations \citep[][]{Charbonneau_2000, ground_base_transit3, ground_base_transit}.
However, the Kepler mission marked a pivotal shift through four years of continuous space-based monitoring of more than 156,000 stars \citep{kepler,Stumpe_2012}. 
This space mission produced an immense volume of high-precision photometric data capable of detecting small planets beyond the sensitivity limits of ground-based observations. 
The sheer scale of the dataset necessitated the development of automated tools for efficient transit signal detection.
Similarly, large-scale survey projects like K2 \citep{Howell_2014} and TESS \citep{tess} have also amassed vast observational datasets, further accelerating the development of automated methods \citep{astronet-k2, tess_1}.

Automated exoplanet detection pipelines generally involve several key steps for identifying transit-like signals.
First, light curves are preprocessed to remove significant stellar activity and variability trends (e.g., \citealt{Stumpe_2012, Smith_2012}). 
Subsequently, the Wavelet-based Adaptive-matched Filter (WAF) \citep{Jenkins2002} or the Box-fitting Least Squares (BLS) \citep{bls} is utilized to recognize periodic signals within the light curves. 
After that, Threshold Crossing Events (TCEs) are identified by applying appropriate thresholds to filter periodic signals detected by WAF or BLS (e.g., \citealt{TCEs_2,KDPS, bls}).
Until now, numerous improvements have been developed in BLS to enhance the speed or precision \citep{bls_imp_4, bls_imp_3, bls_imp_2, optical_sampling}.
For example, \cite{fbls} integrates the Fast Folding Algorithm \citep{ffa} with BLS, significantly accelerating the search speed. 
Meanwhile, the Transit Least Squares (TLS) \citep{TLS} achieves higher precision by adopting the physical limb-darkening model \citep{limbdarkening} to fit the observed data.
However, this increased precision comes at the cost of significantly higher computational complexity.
To address this, the GPU-accelerated Transit Least Squares (GTLS; Q. Hu et al. 2026, in preparation) leverages GPU parallel computing technology to accelerate the TLS computational process.

Next, machine‑learning (ML) screens TCEs for transit candidates, removing false positives caused by instrument noise, eclipsing binaries and stellar variability.
Prominent ML auto-vetting methods include AstroNet \citep{astronet}, Autovetter \citep{autovetter}, and Robovetter \citep{robotvetter}.
In particular, AstroNet is based on the Convolutional Neural Network (CNN, \citealt{cnn_1, cnn_2, cnn_3}), which successfully identified the unknown exoplanets Kepler-80g and Kepler-90i \citep{astronet}. 
The updated versions of AstroNet have been deployed in other surveys, including K2 \citep{astronet-k2}, TESS \citep{tess_1, tess_2, tess_3, tess_4}, WASP \citep{wasp}, and NGTS \citep{ngts}.

Meanwhile, multiple fast automated exoplanet‑detection pipelines skipping least‑squares TCE‑finding methods have advanced greatly \citep{undtrending_1, undrending_2, Chintarungruangchai_2019, Cui_2022, malik}.
Recently, \cite{gpfc_1} developed the GPU Phase Folding and Convolutional Neural Network (GPFC) system, which accelerates light-curve folding to second-level timescales and yields an approximately 7\% performance gain in detecting shallow, ultra-short-period transits relative to BLS.
Using this system, \cite{gpfc_2} identified five previously undetected ultrashort-period exoplanets, including Kepler-158d, Kepler-963c, Kepler-879c, Kepler-1489c, and KOI-4978.02.
Moreover, the GPU-Fold developed in \cite{gpfc_1} makes it possible for large-scale neural networks to quickly search for transits.
Among non‑least‑squares methods, some use light‑curve folding, while others are less sensitive to shallow transits \citep{undtrending_1, Cui_2022}.

Driven by the fast‑growing exoplanet catalogue, research has shifted from single‑system analysis to population‑based statistical investigations (\citealt{population_1, population_2, population_5,population_3, population_4, population_6, population_7}).
On the other hand, small rocky exoplanets ($\leq 1.25\,R_\oplus$) located in the terrestrial planet region of our solar system (from Mercury to Mars), including true Earth analogs (or Earth 2.0s)—Earth-sized planets receiving Earth-like irradiation around solar-type stars (F5V-K5V) remain undetected \citep{kepler22b, HTE_2, et2_2024}. 
To date, only a few habitable super-Earths have been discovered within these period ranges, such as Kepler-442b, Kepler-62e, Kepler-283c, Kepler-1638b, Kepler-1652b, Kepler-22b, Kepler-443b, Kepler-1701b, Kepler-1606b, Kepler-1540b, Kepler-174b, HD 40307g, and HD 206520c according to the \href{https://phl.upr.edu/hwc}{Planetary Habitability Laboratory}. 
These habitable super-Earths have masses between 2-10 $M_{\oplus}$ or radii between 1.25-2.5 $R_{\oplus}$. 
Future space missions such as the \textit{PLATO} mission \citep{plato1, plato2} and the \textit{Earth 2.0} (\textit{ET}) mission \citep{et2, et1, et_new_1, et2_2024,et_new_2} adopt wide-field, ultra-high-precision photometry to expand the known sample of terrestrial exoplanets and aim to discover Earth 2.0s while precisely measuring their occurrence rates.

Importantly, besides the gradual improvement in telescope performance, the algorithm for detecting weak terrestrial transit signals from extensive photometric data is equally crucial.
Detecting these exoplanets is particularly challenging because of their long orbital periods and inherently low Signal-to-Noise Ratios (low-SNR), making them difficult to identify in large astronomical datasets.

In conventional exoplanet detection pipelines, classical search algorithms such as BLS and TLS construct periodograms by evaluating the consistency between phase-folded light curves and predefined transit templates over a grid of trial periods. 
These methods remain powerful and widely used, but their sensitivity can decrease in the long-period, low-SNR regime, where only a few transit events are available and the statistical distinction between genuine shallow transits and noise-induced dips becomes increasingly fragile \citep{bls}.

The primary motivation for developing \textbf{DE}tection in phase-folded \textbf{L}ight curves with c\textbf{O}ntrastive \textbf{S}coring (\textbf{DELOS}) is to improve the detectability of terrestrial-sized planets.
The central challenge addressed by DELOS is not merely the application of machine learning to transit detection but the recovery of rare, shallow, long-period terrestrial-sized signals for which only a small number of transit events are available, and conventional periodogram-based statistics become increasingly fragile under low-SNR and correlated-noise conditions.
In the Solar System, the orbital periods of terrestrial planets span a wide range, from $87.97$ days (Mercury) to $686.98$ days (Mars). 
Owing to this broad period distribution, it is impractical for a single system to efficiently cover the entire range while maintaining high sensitivity to shallow transit signals, as different orbital periods require different binning strategies during data analysis. 
Specifically, excessively fine binning introduces unnecessary computational overhead for short-period searches, whereas overly coarse binning compresses long-period signals into too few bins, significantly attenuating or even obscuring them.
To balance detection sensitivity and computational efficiency, we therefore restrict the initial search window to orbital periods of 100–150 days, targeting Intermediate-to-Long-period (ITL) transit signals. 
This period range serves as a practical first step toward Earth-analog searches: it is long enough to enter the low-SNR, few-transit regime, yet short enough to provide a sufficient number of observable transits for robust method development and validation. 
Future extensions toward $\sim$1-year periods will be necessary to fully address true Earth analogs around Sun-like stars.
The framework is flexible and can be adapted to other period ranges once the algorithm is fully developed.

With DELOS, we significantly improved the sensitivity in detecting ITL low-SNR transit signals. 
We demonstrated that DELOS has higher detection performance and faster speed compared to BLS and TLS.
DELOS also identified several dozen high-scoring transit-like signals in the Kepler data, which are used here only to demonstrate the search capability of DELOS and are not claimed as validated planet candidates. DELOS can be further applied to K2 and TESS, as well as future  PLATO and ET missions.

This paper is organized as follows.
Section~\ref{sec:2} introduces the fundamental principles of DELOS and the complete workflow of the proposed detection system.
Section~\ref{sec:3} describes the data preparation process, including Kepler light-curve preprocessing, simulated data generation, binning strategy, and construction of the DELOS training set.
Section~\ref{sec:4} presents comparative experiments between DELOS and commonly used transit detection algorithms.
Section~\ref{sec:5} applies DELOS to real Kepler mission data, presents high-scoring candidates identified by the model, and evaluates its robustness under both white-noise and red-noise conditions.
Section~\ref{sec:6} discusses the limitations and future improvements of DELOS, with an emphasis on enhancing cotrending and detrending performance to further improve the detection upper bound, as well as comparing its computational efficiency with current fast detection methods.
Finally, Section~\ref{sec:7} concludes the paper by summarizing the advantages of DELOS and discussing potential directions for future research.

\begin{figure}
    \centering
    \includegraphics[width=0.40\textwidth]{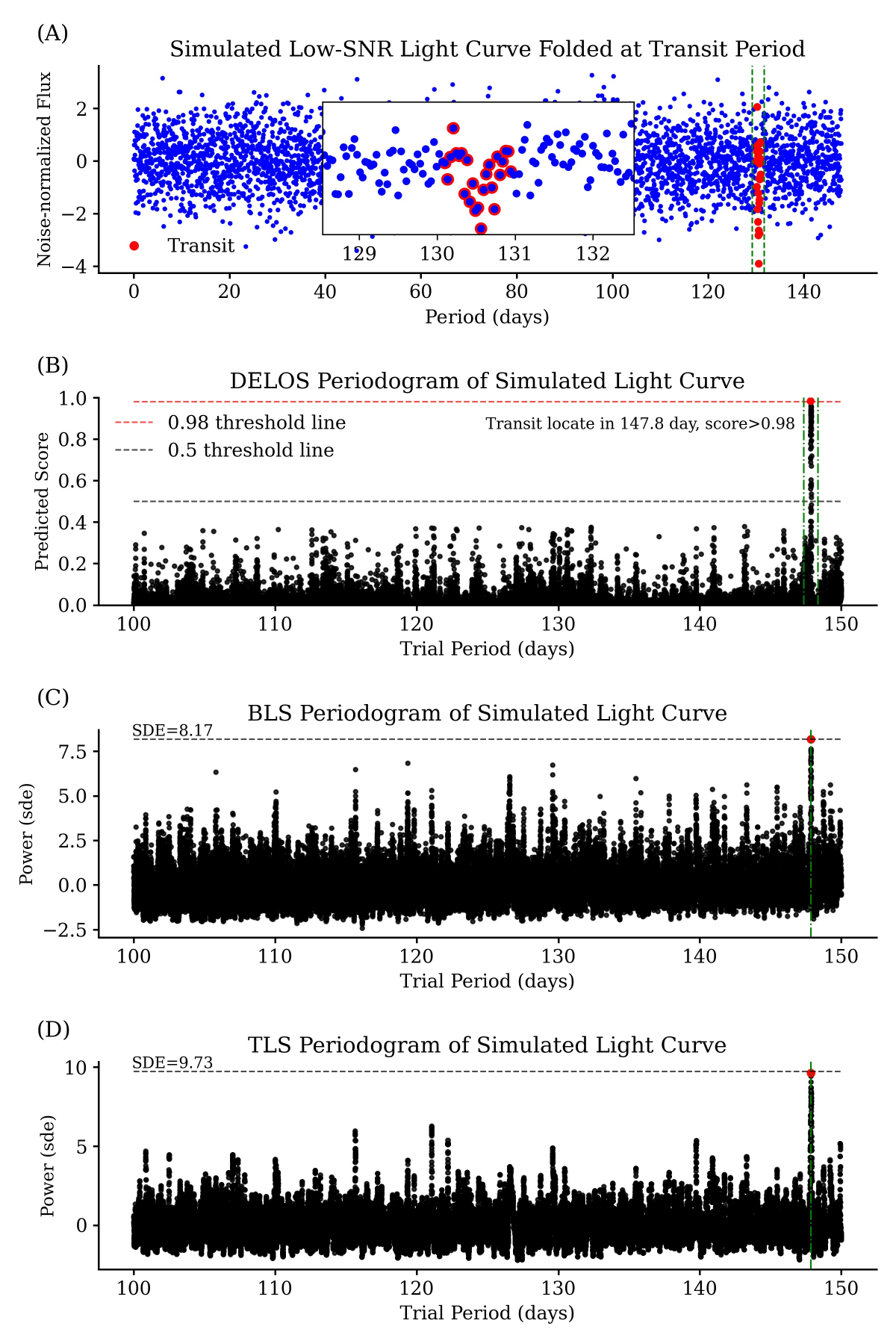}
    \caption{Comparison of the periodograms generated by DELOS, BLS, and TLS on a simulated low-SNR, noise-normalized light curve, each algorithm using the same trial period sampling rate, $N_{\text{sample}} = 58{,}400$. The top panel shows the light curve folded at the correct 147.8-day transit period, highlighting the transit event with a localized zoom. The three panels below sequentially present the periodogram results from DELOS, BLS, and TLS for this light curve: the DELOS score at the transit period exceeds 0.98; the BLS Signal Detection Efficiency (SDE) is 8.17; and the TLS SDE is 9.73.}
    \label{fig:1_figure}
\end{figure}

\section{Methods}\label{sec:2}
\subsection{Overview of the DELOS Method}\label{sec:2.1}

DELOS is a contrastive learning system  designed for detecting ITL, low-SNR transit signals in high precision photometric data. 
This detection system identifies transit signals in light curves through three key stages. 
First, the raw light curve undergoes preprocessing to remove the long-term stellar activity trend and iteratively eliminate outliers. 
The preprocessed light curve is subsequently folded over a range of trial periods using the GPU-Fold algorithm, followed by noise normalization to ensure consistent noise levels across all folded light curves. 
Finally, DELOS assigns a 0–1 transit-likeness score to each normalized light curve, producing a score periodogram in which peaks indicate trial periods with stronger transit-like signatures.
This score serves primarily for the ranking of candidate periods. Unless otherwise specified, it shall not be regarded as a rigorously calibrated astrophysical probability.
In summary, DELOS performs an autonomous transit search without requiring prior period estimates or TCEs from external detection algorithms.

\begin{figure}
    \centering
    \includegraphics[width=0.8\linewidth]{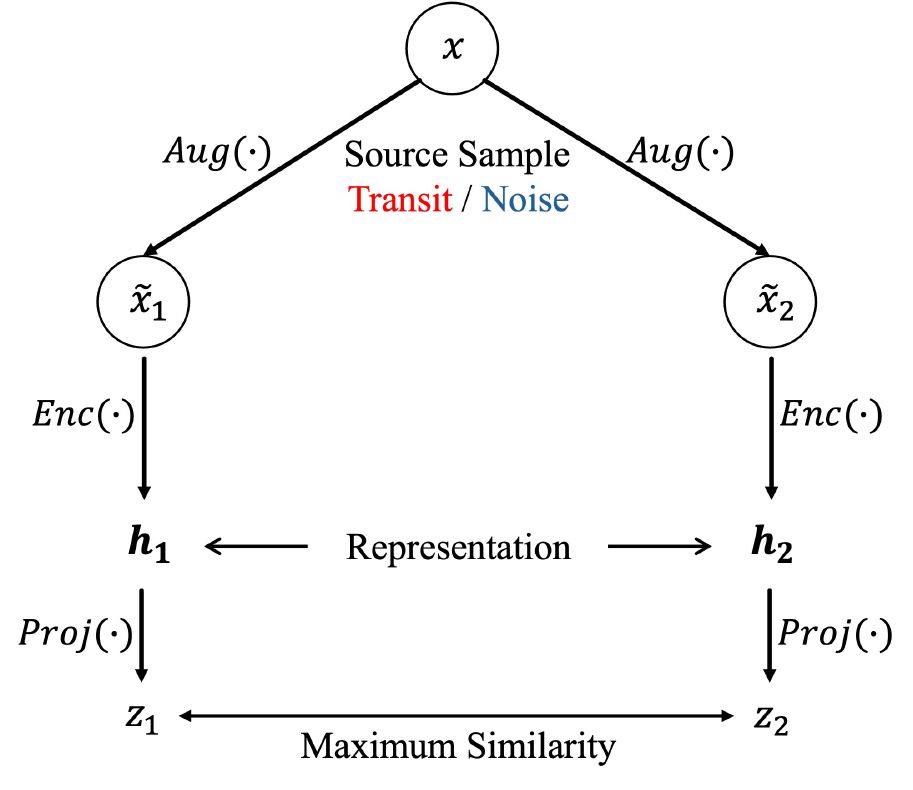}
    \caption{Conceptual diagram of DELOS consisting of four core components: the data augmentation \textit{Aug($\cdot$)}, the encoder network \textit{Enc($\cdot$)}, the projection network \textit{Proj($\cdot$)}, and the contrastive loss. Their relationship can be formulated as $z = \text{\textit{Proj}}(\text{\textit{Enc}}(\text{\textit{Aug}}(x)))$, where the source sample $x$ can represent either transit or noise. The contrastive loss is employed to maximize the similarity between two augmented representations \( z_1 \) and \( z_2 \) derived from the same sample.}
    \label{fig:3figure}
\end{figure}

For the DELOS system, detection sensitivity is strongly influenced by how densely the trial periods are sampled. 
An appropriately chosen sampling frequency is essential to maintain high sensitivity while avoiding excessive computational costs.
We adopt the frequency-grid sampling prescription discussed by \cite{sampleing_formule}, with an oversampling factor ($N_o$).
\begin{equation}
N_{\text{sample}} = \frac{N_o T}{L}
\label{eq:sampling}
\end{equation}
Here, $N_o$ is typically set to 2, $T$ is the total observed time window, and $L$ is the expected transit duration.

The Kepler mission observed 17 quarters of data from 2009 to 2013, yielding a total observational baseline of four years, i.e., $T = 4 \times 365$ days. 
According to the Kepler Object of Interest (KOI) archive \citep{robotvetter}, the transit duration of all known ITL exoplanets (candidates and confirmed planets) falls within 0.05 to 0.93 days.
To ensure optimal sensitivity, we conservatively adopt $L = 0.05$ days as the expected transit duration.
Substituting these parameters into Equation~\ref{eq:sampling} yields a sample frequency of the trial period $N_{\text{sample}} =  \frac{2 \times 4 \times 365}{0.05} = 58,400$.
We uniformly sampled 58,400 trial periods from 100 to 150 days, with an interval of $8.56 \times 10^{-4}$ days between adjacent trial periods, ensuring a sampling density sufficient to detect ITL transit signals.
Next, we employ the GPU-Fold algorithm of \cite{gpfc_1} to fold the preprocessed light curve. 
To avoid producing empty values due to the GPU-Fold mechanism \citep{gpfc_1}, we first fold the preprocessed light curve into 8192 phase bins using the GPU-Fold algorithm, then rebin it to 4096 points for subsequent analysis.
According to the KOI archive, the ratio of transit duration to orbital period for ITL transits ranges from $0.8 \times 10^{-3}$ to $5.5 \times 10^{-3}$. 
Under our binning strategy, this corresponds to approximately 3 to 23 in-transit data points in the binned data, a range sufficient for the model to accurately capture the transit signals.

\begin{figure*}
    \centering
    \includegraphics[width=0.85\linewidth]{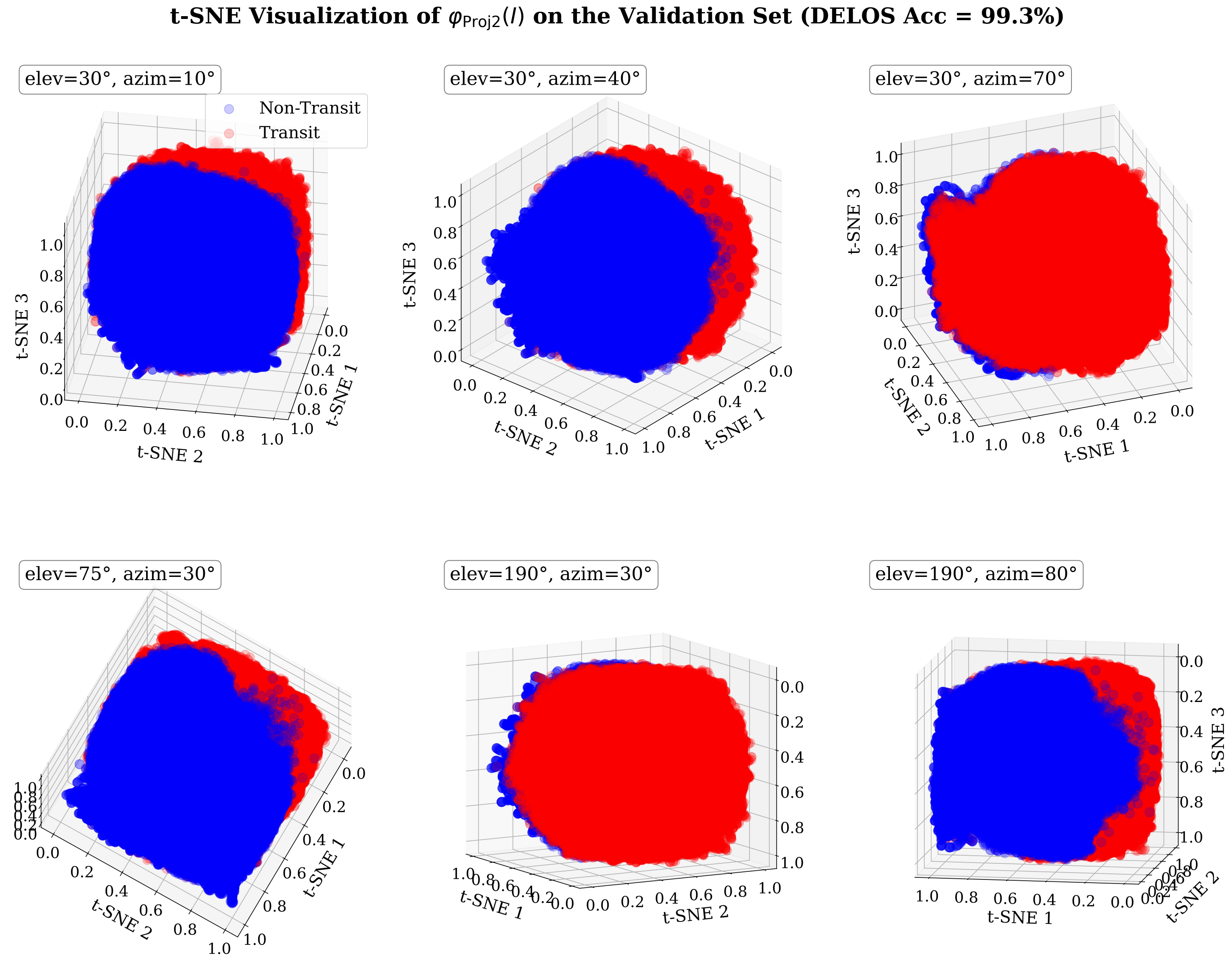}
    \caption{
    Multi-view 3D t-SNE visualization of the projection-head representation for the validation set, shown from six combinations of viewing angles.
    Here, \(I\) denotes the input folded light curve, and \(\varphi_{\mathrm{Proj2}}(I)\) denotes the representation output by the second layer of the projection network.
    Blue points denote no-transit samples, and red points denote transit samples.
    The multiple viewpoints provide a more complete geometric inspection of the embedding distribution and class-wise structure in the learned feature space.
    The consistent clustering patterns across views indicate that DELOS learns high-quality, compact distributional representations, achieving an overall validation accuracy of $99.3\%$.}
    \label{fig:t-SNE}
\end{figure*}

Moreover, the standard deviation of the noise in the binned data is effectively reduced by binning over multiple folded photometric data. 
The degree of noise reduction is approximately proportional to $\sqrt{N}$, where $N$ denotes the number of data points averaged within each bin.
It can be calculated as \( N = \frac{N_{\text{raw}}}{N_{\text{binned}}} \),  where \( N_{\text{raw}} \in [40{,}000,\,70{,}000] \) denotes the number of raw Kepler data points,  and \( N_{\text{binned}} = 4096 \) is the binned number.
Consequently, $1 / \sqrt{N}$ is between 0.24 and 0.32, indicating that the noise is reduced to 24-32\% of its original level. 
Accordingly, the transit SNR is improved, enabling DELOS to detect low-SNR transit signals more effectively.

However, different stellar brightness and variations in $N$ across different stellar light curves lead to differences in the resulting noise levels.
To address this, we normalized the noise distribution of the 58,400 phased light curves to follow $\mathcal{N}(0,1)$, similar to the noise normalization in \cite{gpfc_1}.
Specifically, we first calculated the mean ($u_{\text{binned}}$) and standard deviation ($\sigma_{\text{binned}}$) of the binned light curve, then normalized all $N_{\text{binned}}$ data points by subtracting $u_{\text{binned}}$ and dividing by $\sigma_{\text{binned}}$.
The procedure was then repeated for all 58,400 binned light curves at different trial periods, ensuring that the scores assigned by DELOS remained comparable across light curves with varying noise levels.

Finally, we evaluated the 58,400 normalized light curves using DELOS, generating 58,400 scores that form the score periodogram.
By analyzing the peaks in this periodogram, we can determine the locations of possible transit signals, with the peak magnitudes directly reflecting the confidence of the corresponding transit signals. 
To facilitate a clear comparison of the periodogram outputs from DELOS, BLS, and TLS, we present in Figure~\ref{fig:1_figure} a simulated light curve containing a transit signal with a period of 147.8 days and a transit SNR of 6–7.
The corresponding search results obtained using DELOS, BLS, and TLS are also shown.
The contrast between peaks associated with genuine transit signals and those arising from noise appears to be most pronounced for the DELOS results, followed by those obtained with TLS and BLS.

\subsection{Contrastive Learning}\label{sec:2.2}
DELOS is designed to identify ITL, low-SNR transit signals using a contrastive learning–based framework.
An overview of the method is provided below.

The concept of discriminative similarity learning was first proposed by \cite{cl_1} to effectively address the challenges of face verification tasks involving numerous categories with limited samples per class.
This research was inspired by the Siamese network introduced by \cite{cl_2} to solve the verification of signatures. 
Contrastive learning leverages similarity metrics to compare and learn subtle differences between distinct signatures or faces, thereby learning robust and compact representations that enable accurate discrimination.
This approach parallels the challenge of distinguishing ITL low-SNR transit signals from noise, where the goal is to identify subtle patterns in both the transit signals and noisy data to obtain reliable and accurate representations.
Numerous frameworks have been proposed to derive representations based on similarity more efficiently from data in the field of computer science (e.g., \citealt{cl_list1, cllist_2, cllist_3, cllist_4, cllist_5}). In particular, \cite{CLS} introduced and validated the SimCLR framework, which is aimed at acquiring high-quality and generalizable representations.

The design of most contrastive learning models inevitably consists of four core components: the data augmentation module, the encoder network, the projection network, and the contrastive loss.
The conceptual design of DELOS is shown in Figure~\ref{fig:3figure}.
DELOS employs a supervised contrastive loss to maximize the similarity between representations of different augmented views generated from the same initial sample.
These representations are obtained by sequentially encoding the augmented views through the encoder network and the projection network.
The whole process facilitates DELOS in robustly learning high-quality, compact distributional representations of transit signals and noise in the embedding space.
The detailed roles of the data augmentation module, encoder network, and projection network in DELOS are described in Appendix \ref{app:contrastive_components};
here, we focus on how the supervised contrastive objective separates transit and noise representations.

DELOS uses a supervised contrastive learning loss to optimize the encoder and projection network parameters; this optimization in turn guides the learning direction of the model.
To illustrate the details of supervised contrastive learning loss, we suppose that there are \(N\) initial samples \(\{x_k\}_{k=1 \dots N}\) and their corresponding labels \(\{y_k\}_{k=1 \dots N}\), where \(y_k\) indicates whether \(x_k\) corresponds to transit or noise.
For simplicity, we assume that each initial sample \(x_k\) generates only two augmented views, \(\tilde{x}_{2k-1}\) and \(\tilde{x}_{2k}\) through the data augmentation module \( \textit{Aug}(\cdot) \) along with their labels \(\tilde{y}_{2k-1}\) and \(\tilde{y}_{2k}\), where \(y_k = \tilde{y}_{2k-1} = \tilde{y}_{2k}\), i.e., each transit or noise sample includes only two views here.
A total of \(2N\) views and their corresponding labels are generated for model training (DELOS in practice generates \(100N\) views), denoted as \(\{\tilde{x}_i, \tilde{y}_i\}_{i \in \mathcal{I} = \{1 \dots 2N\}}\), where \(\mathcal{I}\) denotes the index set of all augmented views.
Each augmented view \(\tilde{x}_i\) is then mapped to the projection representation \(z_i = \text{\textit{Proj}}(\text{\textit{Enc}}(\tilde{x}_i))\), which is used in the supervised contrastive loss.

Among these \(2N\) views, 
\(A(i)\) denotes the index set of all augmented views except \(\tilde{x}_i\), i.e., \(A(i) = \mathcal{I} \setminus \{i\}\), and thus \( A(i) \) contains \(2N-1\) elements.
Although Figure~\ref{fig:3figure} illustrates two augmented representations derived from the same source sample, the supervised contrastive loss \citep{supercontrastive} used below treats all augmented views with the same label as positive samples.
\begin{equation}
    \mathcal{L}^{\text{sup}}
    = - \sum_{i \in \mathcal{I}} \frac{1}{|P(i)|}
    \sum_{p \in P(i)}
    \log
    \frac{\exp \left(z_i \cdot z_p / \tau\right)}
    {\sum_{a \in A(i)} \exp \left(z_i \cdot z_a / \tau\right)} ,
    \label{eq:scl}
\end{equation}
Here, \(P(i)\) denotes the index set of all other augmented views among the \(2N\) views that belong to the same label as \(\tilde{x}_i\), i.e., \( P(i) \equiv \{ p \in A(i) : \tilde{y}_p = \tilde{y}_i \} \), and \( |P(i)| \) is the cardinality of \( P(i) \).
The dot symbol \( \cdot \) represents the inner product, which is used to quantify the similarity between \(z_i\) and \(z_p\).
The parameter \(\tau\) is a scalar temperature constant that controls the scaling of similarity scores in contrastive learning; 
it influences the sharpness of the resulting distribution and thus affects optimization dynamics and generalization. 
\(\tau\) is often chosen as a positive real number, e.g., 0.07.

In this supervised contrastive loss shown in Equation~\ref{eq:scl}, the numerator \(\exp \left(z_i \cdot z_p / \tau\right)\) quantitatively calculates the similarity between $z_i$ (encoded \(\tilde{x}_i\)) and all other augmented views that share the same label, while the denominator \(\sum_{a \in A(i)} \exp \left(z_i \cdot z_a / \tau\right)\) quantitatively calculates the similarity between $z_i$ and the remaining \(2N-1\) augmented views.

As illustrated in Figure~\ref{fig:t-SNE}, DELOS pulls representations of same-label samples closer together in the embedding space while pushing representations of different-label samples farther apart.
Finally, DELOS learns both the similarity among samples sharing the same label and the discriminative differences between samples with different labels.

\subsection{Overall Framework of DELOS}\label{sec:2.3}
\begin{figure}
    \centering
    \includegraphics[width=1.0\linewidth]{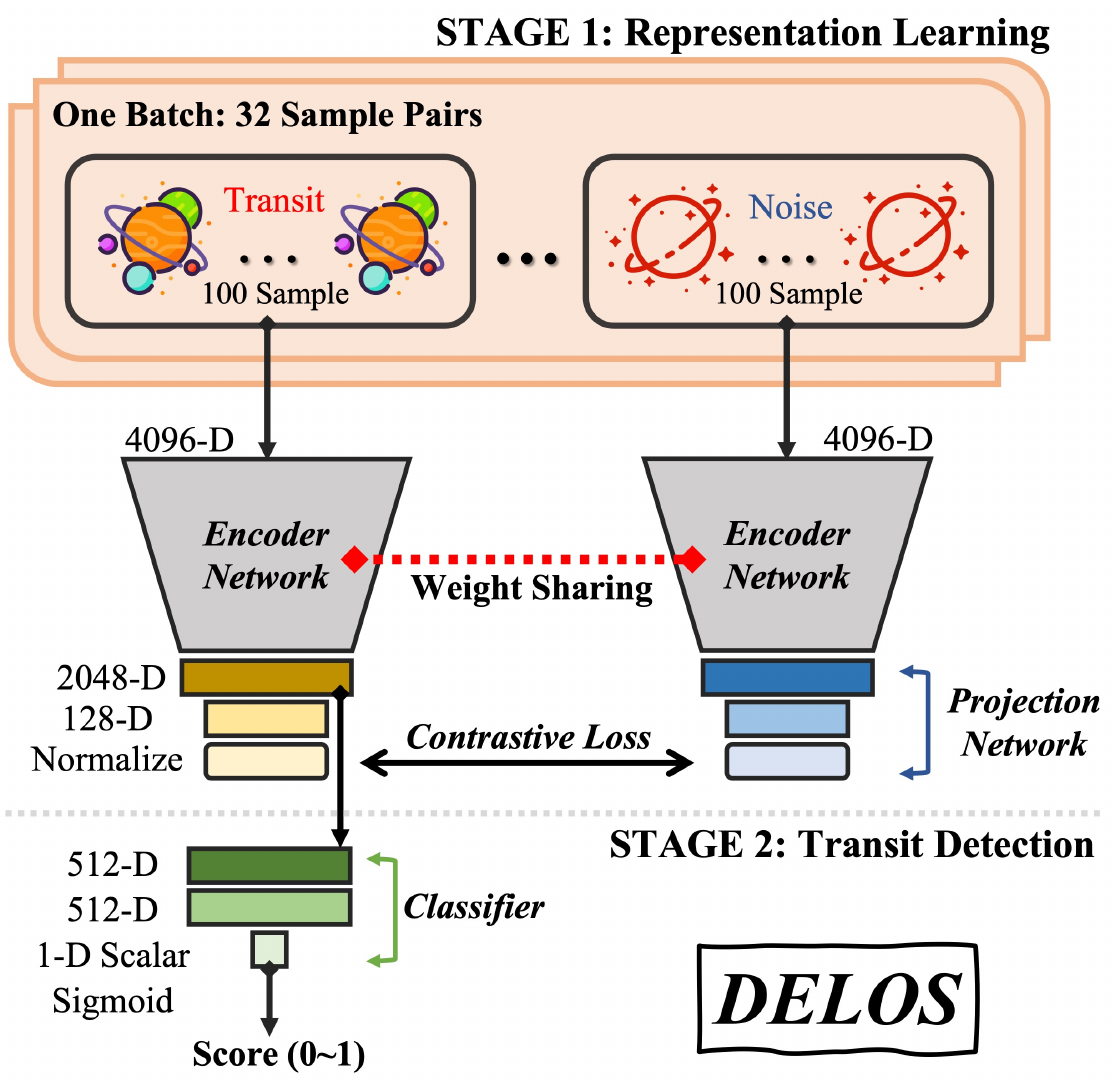}
    \caption{Schematic illustration of the DELOS framework. The framework consists of two stages: \textit{Representation Learning} and \textit{Transit Detection}.
    In the representation learning stage, augmented pairs generated from real transit and noise light-curve samples are used to optimize the encoder and projection network.
    Labels such as 4096-D and 2048-D denote the dimensionality of the corresponding input or latent feature vectors.
    In the transit detection stage, folded light curves are assigned transit-likeness scores.}
    \label{fig:5figure}
\end{figure}

DELOS is built upon a supervised contrastive learning framework tailored for the detection of ITL low-SNR transit signals.
Its high performance is driven by the combined effects of the contrastive learning mechanism and the structured data organization strategy. 
To summarize this organization, Figure~\ref{fig:5figure} presents the two-stage architecture of DELOS, consisting of the representation learning stage and the transit detection stage.

\begin{figure*}
  \vspace{0em}
  \centering
  \captionsetup{skip=1pt}

  \begin{subfigure}[t]{0.48\textwidth}
    \centering
    \includegraphics[width=0.95\linewidth]{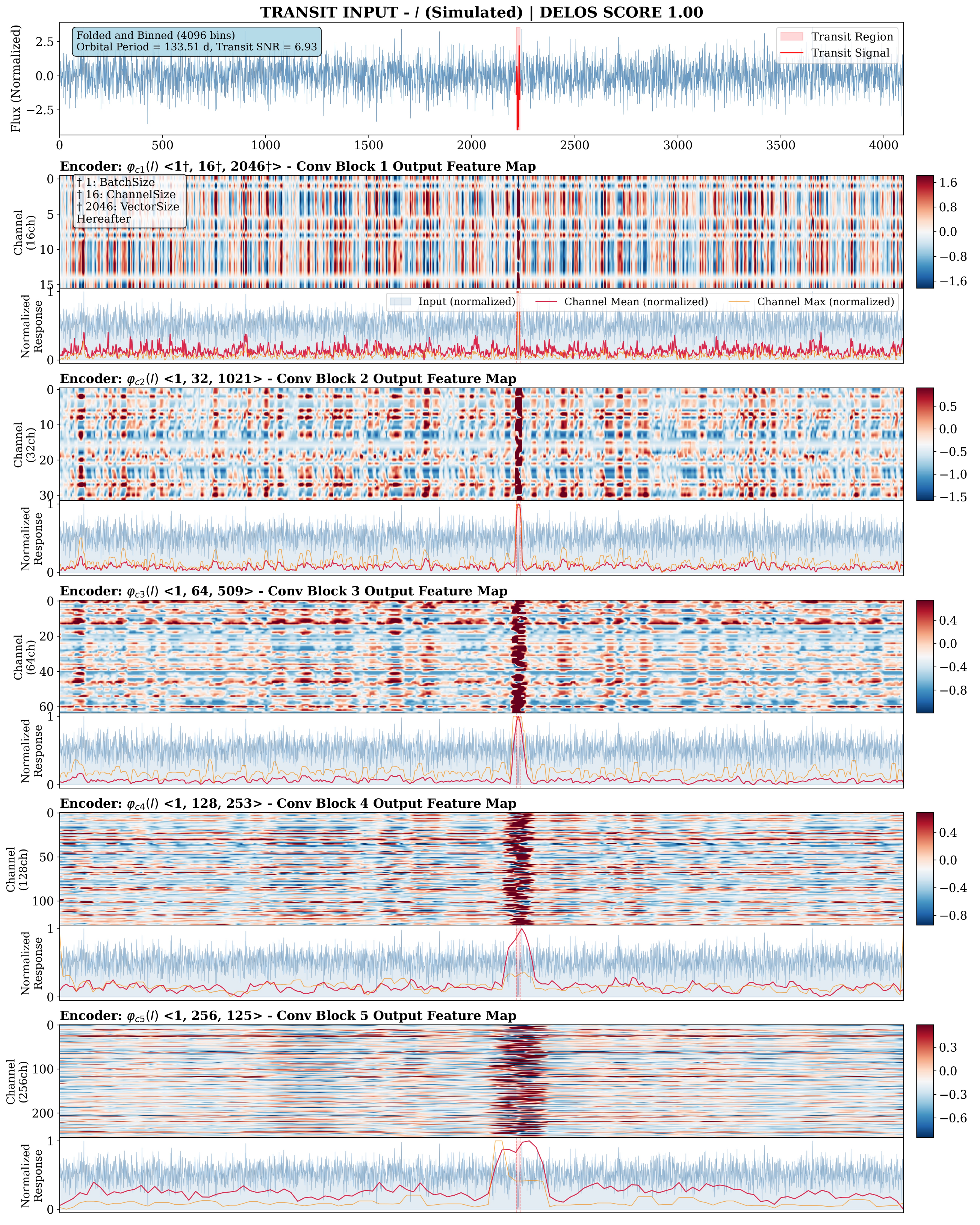}
  \end{subfigure}\hfill
  \begin{subfigure}[t]{0.46\textwidth}
    \centering
    \includegraphics[width=0.95\linewidth]{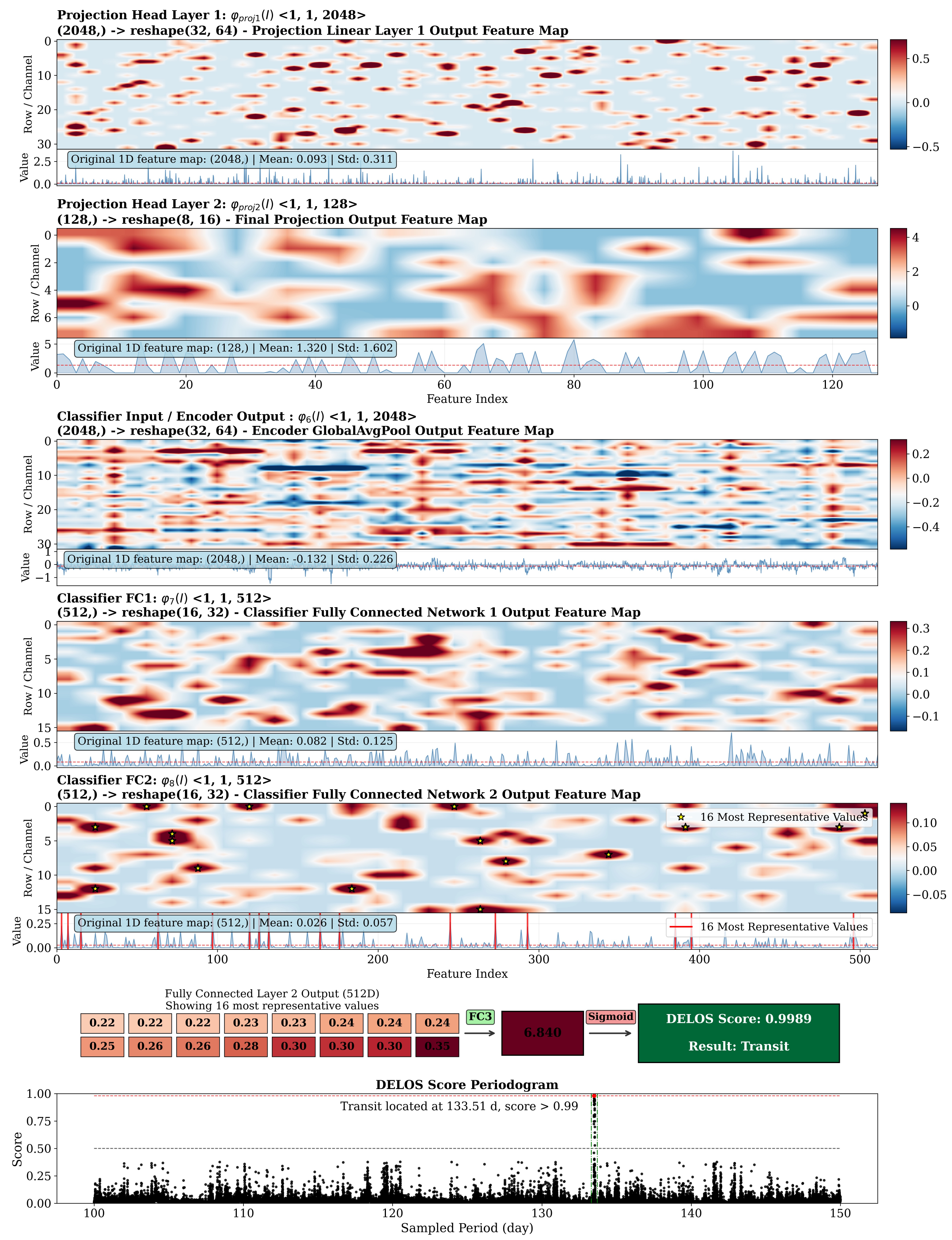}
  \end{subfigure}

  \vspace{0.0em}
  \noindent\begin{tikzpicture}[baseline]
    \draw[dashed] (0,0) -- (\textwidth,0);
  \end{tikzpicture}
  \vspace{0.0em}

  \begin{subfigure}[t]{0.48\textwidth}
    \centering
    \includegraphics[width=0.95\linewidth]{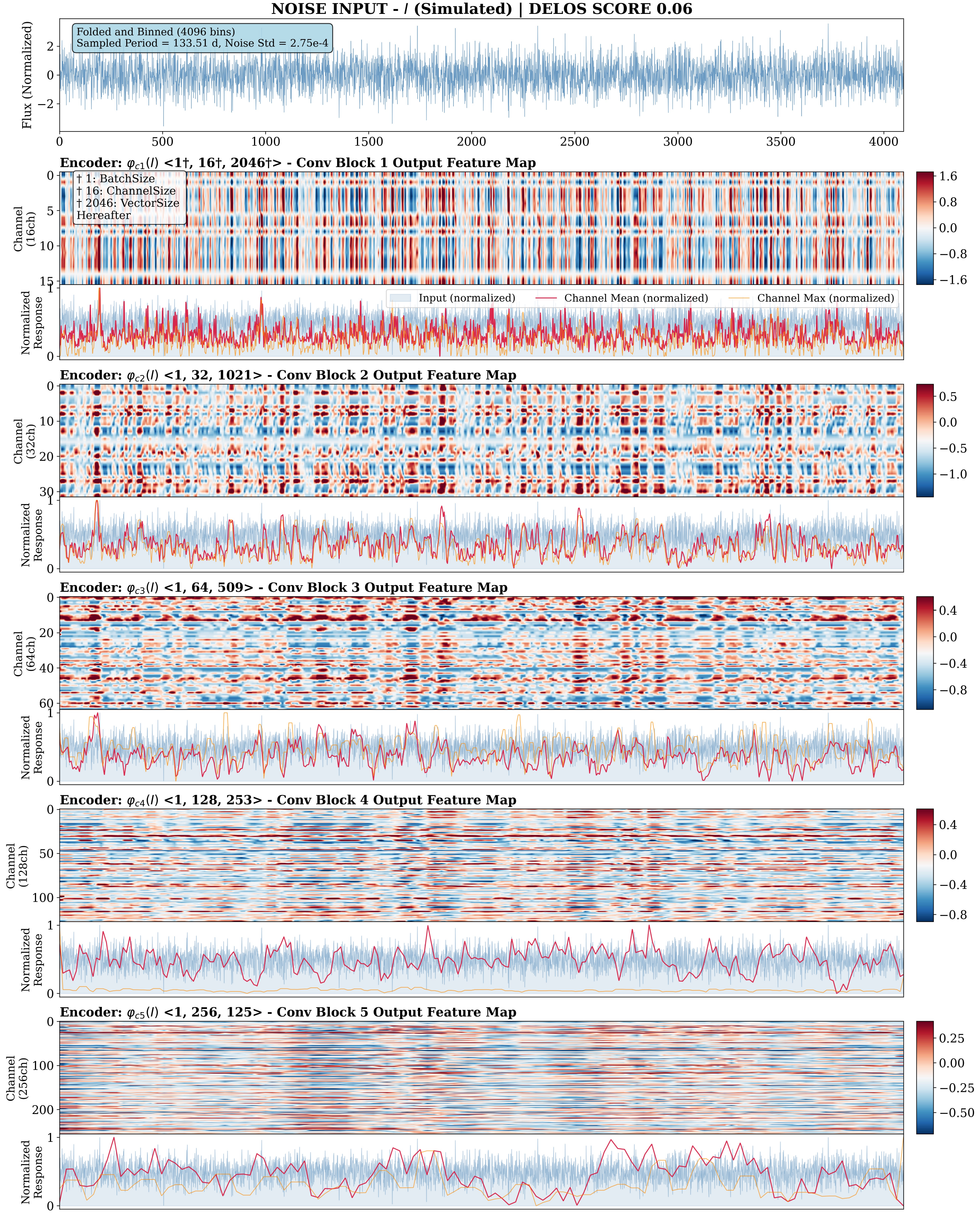}
  \end{subfigure}\hfill
  \begin{subfigure}[t]{0.45\textwidth}
    \centering
    \includegraphics[width=0.95\linewidth]{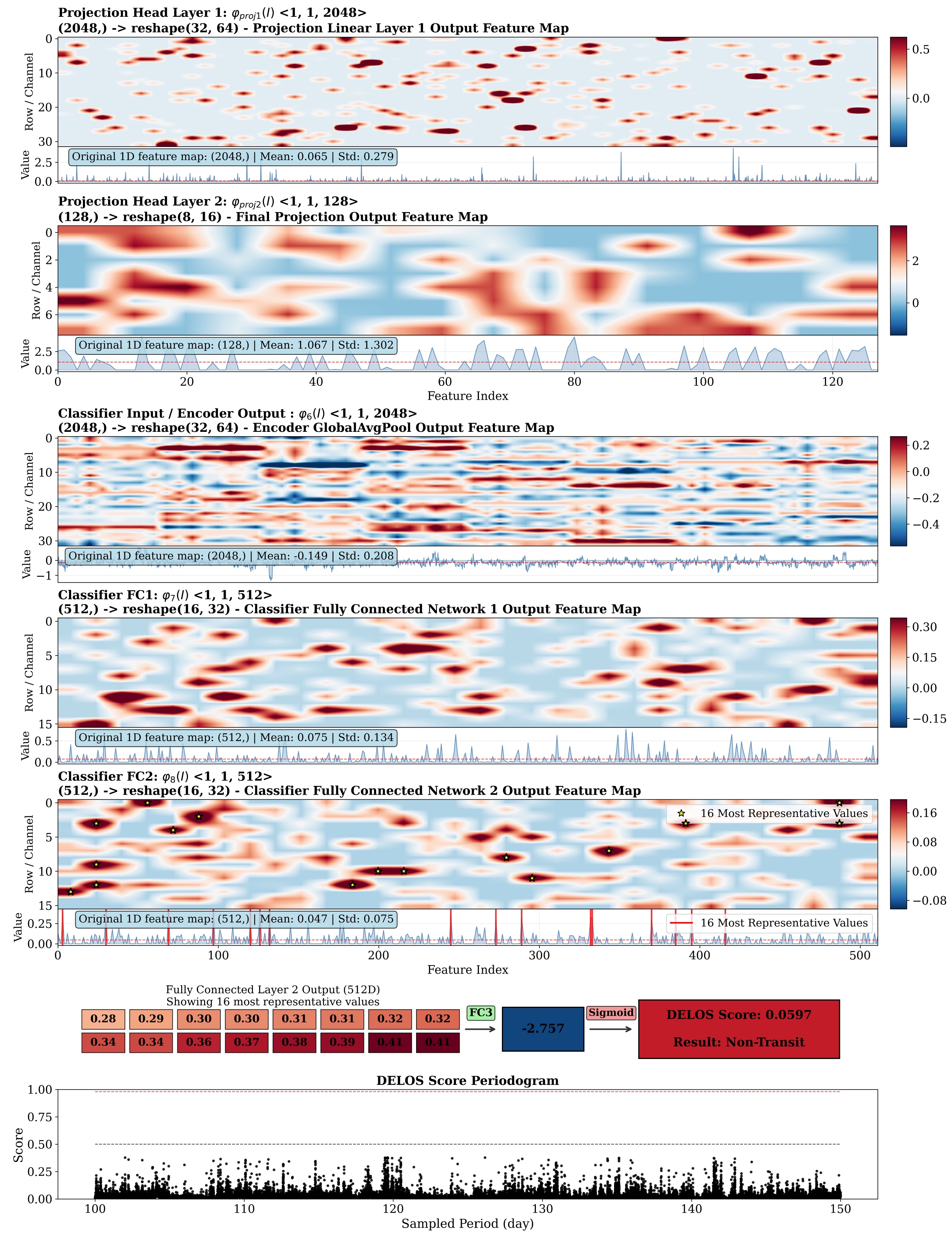}
  \end{subfigure}

    \caption{End-to-end feature evolution and decision process of DELOS for two simulated examples: transit signal (top panel) and noise (bottom panel).
    The left column shows the folded-and-binned input light curve (4096 bins) and the feature maps of the five encoder convolution blocks, \(\varphi_{c_i}(I)\) (\(i=1,\dots,5\)), with channel-statistic traces (e.g., mean/max) summarizing layer-wise responses.
    The right column shows the projection-head outputs (e.g., \(\varphi_{\mathrm{Proj1}}(I)\), \(\varphi_{\mathrm{Proj2}}(I)\)), the GlobalAveragePooled feature \(\varphi_{6}(I)\), and the classifier activations \(\varphi_{7}(I)\) and \(\varphi_{8}(I)\), together with highlighted high-response units.
    The decision blocks report the final score \(s\in[0,1]\) and predicted class, while the bottom-right subpanel shows the DELOS score periodogram \(s(P)\), whose peak indicates the most likely period.}
    \label{fig:feature_map}
\end{figure*}

DELOS was developed and optimized to resolve the confusion between the representation of shallow transit and noise. 
Specifically, we generated 100 distinct transit views by randomly shifting the transit midpoint time of an initial ITL low-SNR transit source sample.
We define these views as a transit pair.
Similarly, we injected an ITL low-SNR transit signal into four years of light curve data and randomly folded it using incorrect periods to generate 100 distinct noise views, which we define as noise pairs. 
We generated 100,000 distinct transit pairs and 100,000 distinct noise pairs.
With each sample pair containing 100 light curves, the total number of simulated light curves is 20 million.
Afterward, numerous pairs are input into DELOS, and the contrastive loss and classification loss are used to iteratively optimize the network parameters of the Encoder, Projection, and Classification layers through backpropagation.

The implementation settings, network structures, and notation of the intermediate features associated with this framework are provided in Appendix~\ref{app:delos_implementation}, while the main text focuses on the overall training organization and end-to-end decision process.

The end-to-end feature evolution and decision-making process, including the encoder feature maps, projection-head representations, classifier activations, and score periodogram, is illustrated in Figure~\ref{fig:feature_map}. 
The transit sample exhibits progressively concentrated transit-related responses and a clear dominant peak in the score periodogram, whereas the noise sample shows diffuse activations without a significant peak, leading to a low final score.
Although DELOS requires constructing numerous transit and noise pairs during training, for the transit search, we only need to input each light curve into the optimal DELOS model without the need to reconstruct pairs.
This is because DELOS has learned high-quality, compact transit and noise representations, which provide two vital insights:

\begin{itemize}
\setlength{\itemsep}{2pt}
\setlength{\topsep}{0pt}
    \item The significant similarity of numerous ITL low-SNR transits in the embedding space.
    \item A clear gap between ITL low-SNR transits and noise in the embedding space.
\end{itemize}

\section{Data preparation}\label{sec:3}
\subsection{Pre-processing of Light Curve}\label{sec:3.1}
We selected targets satisfying the ITL and low-SNR criteria from the KOI archive and downloaded their \textit{Kepler} Pre-search Data Conditioning Simple Aperture Photometry (PDCSAP) long-cadence light curves according to their \textit{Kepler} Input Catalog (KIC) \citep{KDPS} from the \href{https://archive.stsci.edu/pub/kepler/lightcurves/}{Data Archive} of \textit{Kepler}. The data we downloaded have already undergone the Cotrending process by the \textit{Kepler} Data Science team.
Each light curve contains observation time and photometric measurements, with a time interval of $29.4244$ minutes between data points.
The total duration of the observation is $4$ years, divided into 17 quarters, and the total number of data points ranges between 40,000 and 70,000.
Next, we primarily remove the low-frequency systematic components in the raw light curves, which are mainly attributed to stellar activity,  as part of the Detrending process similar to the approaches in \cite{gpfc_1}.

\begin{figure}[ht!]
    \centering
	\includegraphics[width=0.4\textwidth]{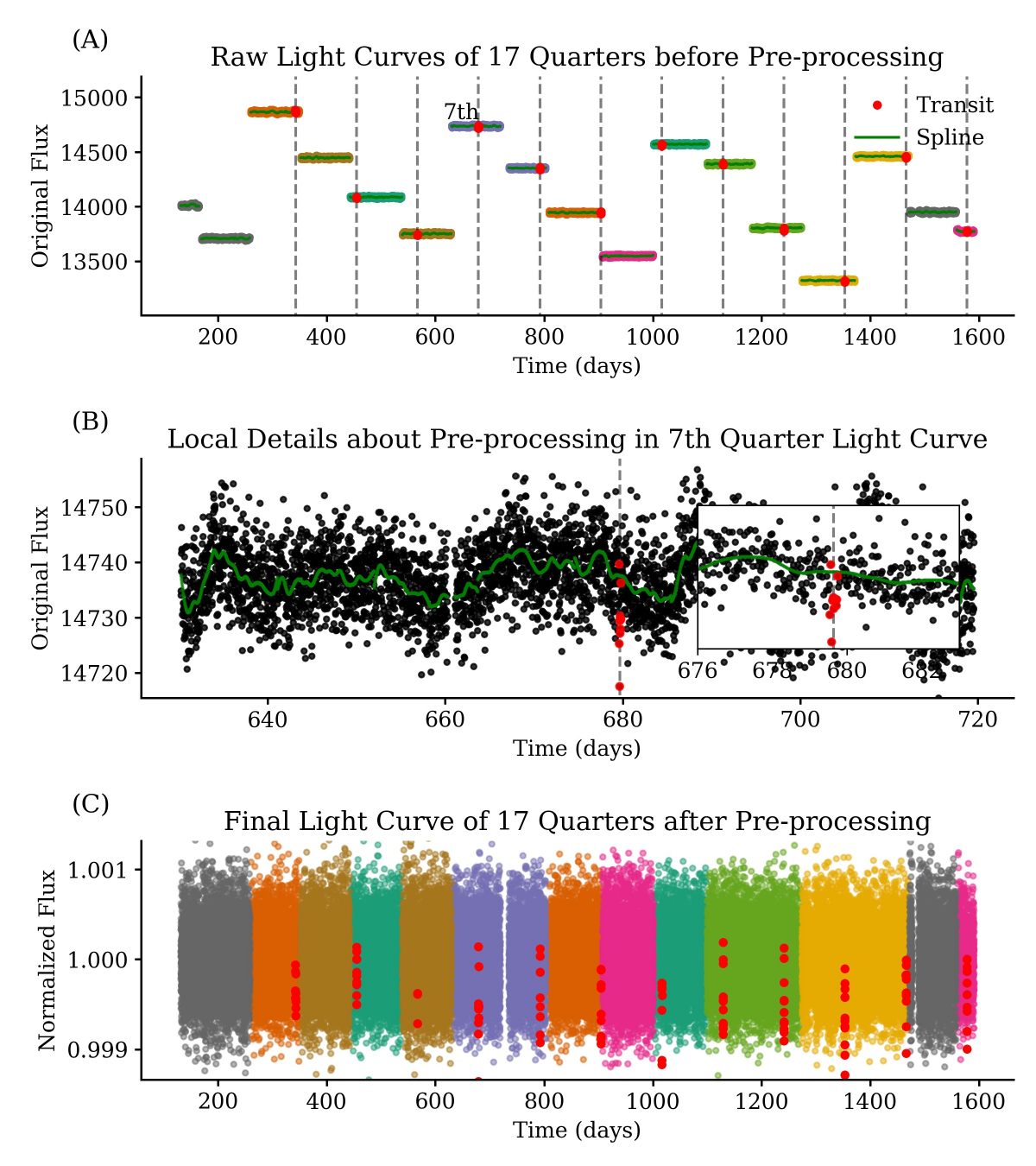}
    \caption{
    Preprocessing of a real \textit{Kepler} light curve and its changes before and after.
    We used a known ITL transit signal KIC 4138008 as the example. The top panel displays the raw PDCSAP light curve across 17 quarters, with the transit events marked in red.
    Dashed lines indicate the timing of each transit event, while the green line represents the best fitted smooth curve.
    The bottom panel presents the final light curve after preprocessing, which is ready for input into the DELOS detection system.}
    \label{fig:2_figure}
\end{figure}

We first masked all known transit signals so that subsequent processing would not alter them.
Next, for each quarter, we split the light curve into multiple segments based on time gaps and discarded the first and last 30 cadences to mitigate systematics associated with the quarterly roll maneuvers or data downlinks of the \textit{Kepler} spacecraft.
We then iteratively fit a smoothing spline curve to each segment to model long-term stellar variability, rejecting per-segment $3\sigma$ outliers.
In practice, the smoothing parameter of the spline model was adaptively tuned using two criteria: 
\begin{itemize}
\setlength{\itemsep}{2pt}
\setlength{\topsep}{0pt}
    \item Residual sign balance: approximately equal counts of data points above and below the spline curve.
    \item RMS conservation: the Root-Mean-Square of the light curve changes minimally before and after detrending.
\end{itemize}

Finally, we divided the PDCSAP light curve by the best-fit spline and concatenated the segments in chronological order.
These steps were repeated for each quarter, and the preprocessed data served as input to the DELOS detection system. 
In addition, when searching for previously unknown exoplanets, we retained downward 3$\sigma$ outliers while directly removing all known transit signals.

An example of the preprocessing applied to a real \textit{Kepler} target is shown in Figure~\ref{fig:2_figure}.
The raw PDCSAP light curve of KIC 4138008, a detailed local view of the fit in the seventh quarter, and the resulting preprocessed light curve are presented.
After the dominant components of stellar activity are effectively removed, the transit signals become clearly visible.
The preprocessed data can then be used by DELOS to validate known candidates or to conduct blind searches for previously unknown intermediate-to-long-period transit signals.

\subsection{Light Curve Simulation}\label{sec:3.2}
\begin{figure}[ht!]
    \centering
    \includegraphics[width=0.9\linewidth]{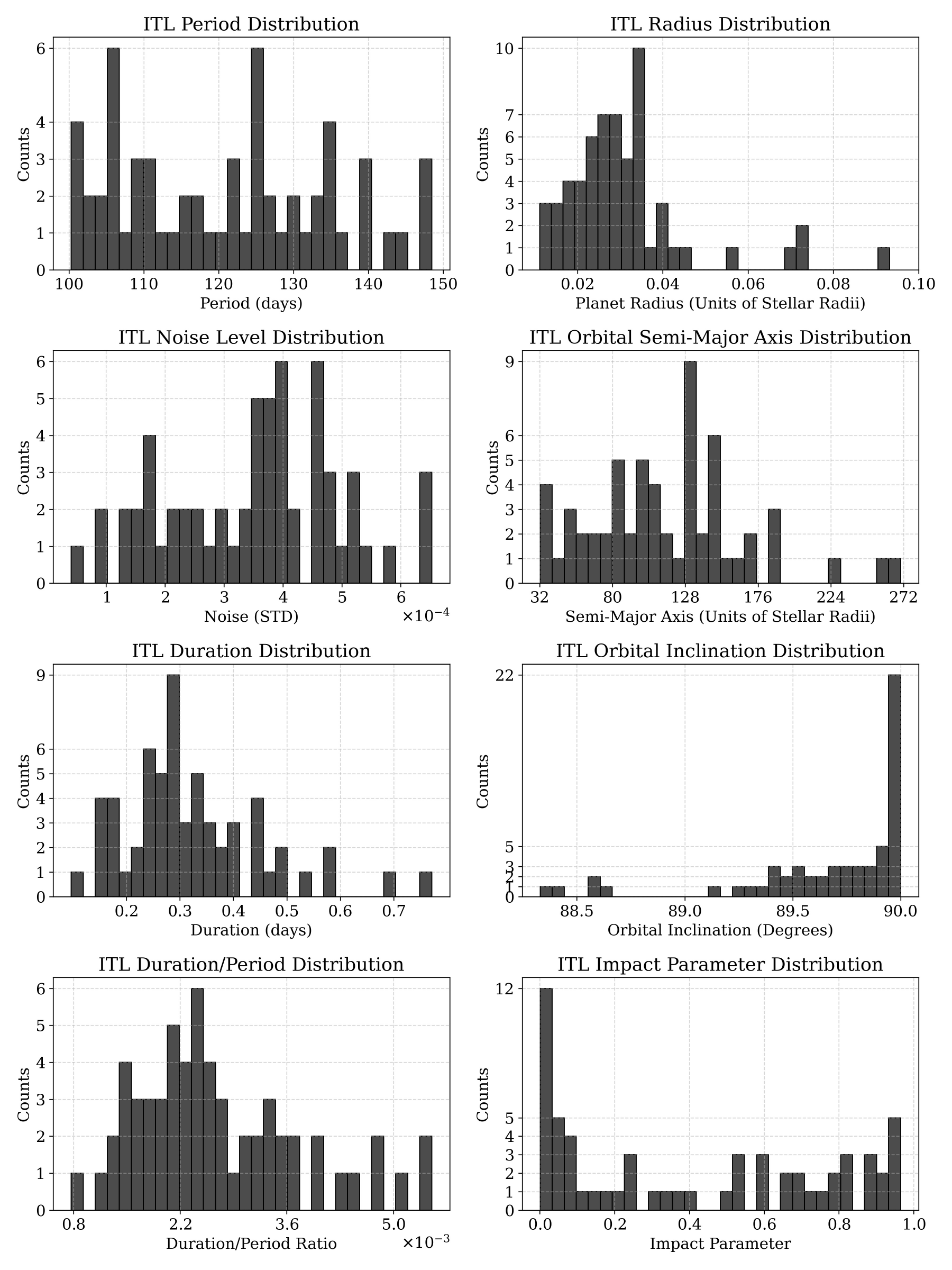}
    \caption{
    Distributions of key intrinsic parameters used to simulate light curves. 
    In practice, the ITL noise level is determined by measuring the $1\sigma$ white noise level in the light curves after Detrending.
    }
    \label{fig:7figure}
\end{figure}

The scarcity of real ITL transit samples poses a significant challenge for effective model training.
To address this limitation, we used \href{https://lkreidberg.github.io/batman/docs/html/index.html}{\texttt{BATMAN}} \citep{batman} to generate numerous simulated light curves based on the quadratic limb-darkening transit model.
Specifically, we selected 60 confirmed ITL transits from the KOI archive and extracted key transit parameters for simulation.
The distributions of these parameters are shown in Figure~\ref{fig:7figure}.
We applied the formula from \cite{snr} to compute the transit signal-to-noise ratio, presented as follows:
\begin{equation}
Transit\ SNR = \frac{d}{\sigma} \sqrt{n \frac{L}{p}}
\label{snr_fomule}
\end{equation}
Here, \( d \) is the transit depth, \( \sigma \) is the uncertainty in the photometric data, assuming identical uncertainty for each data point.
\( n \) is the total number of data points in the light curve, \( L \) is the duration of the transit signal, and \( p \) is its period. Consequently, \( n \frac{L}{p} \) is the number of data points in the light curve associated with transit events.

To achieve the scientific goal of DELOS in detecting ITL, low-SNR transit signals, we simulated the light curve with \( p \in [100, 150] \) days and transit SNR \(\in [6, 15] \), ensuring that each combination of the parameter space (\textit{p}, \textit{transit SNR}) contains equal light curves.
To accelerate the process, we used 50 parallel processes, each assigned to a 1-day interval of \(p\) (e.g., the first process generates data for \(p \in [100,101)\,\mathrm{d}\)), with uniform sampling performed within that interval.
Within each process, the transit SNR is also evenly divided into 9 subintervals with a step size of 1, and the light curves are simulated sequentially within each subinterval to ensure equal sample sizes across different levels of transit SNR.

For instance, when simulating the light curve with transit SNR \( \in [8, 9] \) in the first process, period \(p_0\) is sampled from \(\mathcal{U}(100, 101)\).
Subsequently, transit midpoint \(t_0\), planetary radius \(r_p\), orbital semimajor axis \(a\), orbital inclination \(i\), limb-darkening coefficients \(u_1\) and \(u_2\), and noise level \(\sigma\) are uniformly sampled from the reasonable ranges shown in Figure~\ref{fig:7figure}.
In particular, \(t_0\) is sampled from \(\mathcal{U}(0, p_0)\), \(u_1\) is sampled from \(\mathcal{U}(0.20, 0.62)\), \(u_2\) is sampled from \(\mathcal{U}(-0.05, 0.38)\), and these bounds of \( u_* \) are derived by cross-referencing the values of \(\log g\), \(T_{\text{eff}}\), and \(\log [\text{M/H}]\) for 60 well-characterized stars with the tabulated results provided in the \href{https://cdsarc.cds.unistra.fr/ftp/J/A+A/529/A75/}{Database Website} published by \citet{limb-darkening-model}.

Once the parameter set $\langle$$p_0$, $t_0$, $r_p$, $a$, $i$, $e$(fixed 0), $w$(fixed 90$^\circ$), $u_1$, $u_2$, $\sigma$$\rangle$ is sampled, and the key transit physical parameters must satisfy certain degeneracies, including the following three \citep{bls}:

\begin{itemize}
\setlength{\itemsep}{0pt}
\setlength{\topsep}{0pt}
    \item $p_0$ and $a$ are related by Kepler's third law, which, for circular orbits, is given by:
    \begin{equation}
    p_0 = 2 \pi \sqrt{\frac{a^3}{G M_*}}
    \end{equation}
    Here, $G$ is the gravitational constant, and $M_*$ is the stellar mass; we take $M_* = M_{\odot}$.
    \item The transit duration $T_{\text{dur}}$ is related to the impact parameter $b$ by:
    \begin{equation}
    T_{\text{dur}} = \frac{p_0}{\pi} \sqrt{1 - b^2}
    \end{equation}
    Here, $b$ is the impact parameter normalized by stellar radius, the relationship between $b$ and the orbital inclination $i$ (for circular orbits, $e=0$) is given by:
    \begin{equation}
        b = \cos(i)
    \end{equation}
    \item The semi-major axis $a$ is related to the transit duration $T_{\text{dur}}$ and the impact parameter $b$ by:
    \begin{equation}
    a = \frac{T_{\text{dur}}^2 \cdot G M_*}{\pi^2 (1 - b^2)}
    \end{equation}
\end{itemize}

Finally, \texttt{BATMAN} is used to simulate a pure light curve, and noise with \(\sigma\) standard deviation is added to it to produce the synthetic light curve.
We then calculated the transit SNR value of the synthetic light curve following Equation~\ref{snr_fomule}. If transit SNR \( \notin [8,9]\) or if the duration and duration/period ratio are outside the valid ranges illustrated in Figure~\ref{fig:7figure}, a new parameter set is resampled until a valid synthetic light curve is produced.

\subsection{Synthetic Data Set for DELOS Training}\label{sec:3.3}
To generate 100,000 distinct transit pairs and 100,000 distinct noise pairs for training DELOS, we synthesized light curves resembling those of \textit{Kepler}, featuring an observational timescale of approximately 4 years and a time resolution of 29.4244 minutes. 
So we sampled the number of data points from \( \mathcal{U}(40,000, 70,000) \) and fixed the time interval between each point at 29.4244 minutes.
First, an original 4-year light curve was simulated by the pipeline of Section~\ref{sec:3.2} and folded into the correct transit period. 
Then, the transit midpoint of the folded light curve was randomly shifted 100 times to produce one transit `pair'. Following the definition of a `pair' in the early contrastive learning papers (e.g., \citealt{supercontrastive}), this pair includes 100 transit views here.
Simultaneously, the original light curve was randomly folded into incorrect periods 100 times to produce one noise pair that includes 100 noise views.
This operation effectively covers the vast majority of potential false signals in real transit search scenarios, providing enhanced robustness for the model.
All folded views were binned into 4096 points before being used as DELOS inputs; the choice of this binning number is discussed in Appendix~\ref{app:optimal_binning}.

\begin{figure}[ht!]
    \centering
    \includegraphics[width=0.9\linewidth]{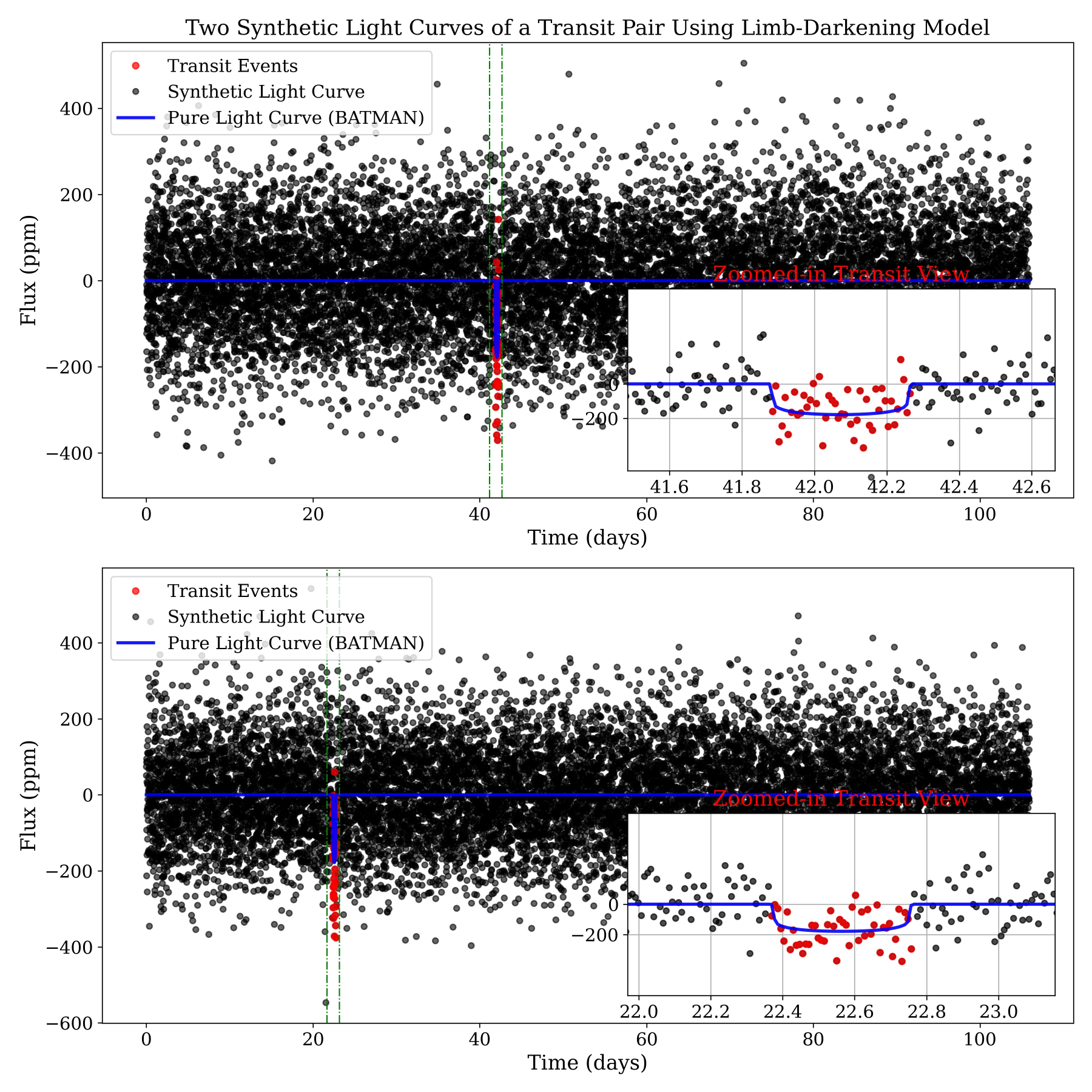}
    \caption{Two synthetic light curves of a transit pair based on the quadratic limb-darkening transit model, and these two light curves differ only in the transit midpoint parameter \(t_0\).
    The blue line represents the pure light curve generated by \texttt{BATMAN}, the black dots indicate the light curve after injecting noise at the $\sigma$-level, the red dots mark the transit events, and the green dashed lines highlight the positions of transit events. Additionally, a zoomed-in view of the transit event is provided in the bottom-right corner of each panel.
}
    \label{fig:8figure}
\end{figure}

By repeating this process for each combination of parameter space (\textit{p}, \textit{transit SNR}), 100,000 transit pairs and 100,000 noise pairs were generated.
For one transit pair characterized by  $p_0$ = 105.91 days, \textit{duration} = 9.19 hours, \textit{transit SNR}= 9.79, \textit{depth} = 178.34 ppm, $\sigma$ = 0.00026, we present two out of 100 transit views in Figure~\ref{fig:8figure}.

After obtaining these data pairs, they were randomly divided into three subsets: 80 percent for the training set, 10 percent for the validation set, and the remaining 10 percent for the test set.
The split was performed at the level of the original simulated source light curve, so that augmented views derived from the same source were never shared between the training, validation, and test sets.
In order to ensure stable training and mitigate issues such as gradient explosion or poor convergence, we used techniques such as Kaiming initialization \citep{kaiming_init} and learning rate warm-up. 
In addition, an early stopping mechanism was implemented to prevent excessive training times and to detect the optimal stopping point.

The runtime information for data generation and the training convergence curves are provided in Appendix~\ref{app:runtime_convergence}. 
After convergence, DELOS achieved its best performance in the 12th epoch with a validation accuracy of 99.3 percent, demonstrating optimal generalization to unseen data.

\section{Comparison}\label{sec:4}
\subsection{Transit-Search Benchmarks: BLS and TLS}\label{sec:4.1}
To evaluate the transit-search performance of DELOS, we conducted comparative experiments with two widely used methods: BLS \citep{bls} and TLS \citep{TLS}.
In brief, BLS is a periodogram-based grid search that evaluates a large set of trial periods together with corresponding transit epochs (phases) and durations using a box-shaped transit model. 
For each trial configuration, the transit depth (and baseline level) is determined analytically by minimizing the squared residuals between the model and the observed flux, and the resulting fit improvement is used to construct the BLS periodogram and identify candidate periods.

Meanwhile, the Signal Residual (SR) statistic is computed in each trial period to obtain the periodogram of SR versus \textit{trial period}.
The standardized detection-strength of the transit signal in the observed light curve is evaluated by calculating the SDE, shown as follows:
\begin{equation}
SDE = \frac{SR_{\text{peak}} - \langle SR \rangle}{\sigma_{SR}}
\label{eq:SDE}
\end{equation}
Here, \( SR_{\text{peak}} \) is the maximum value in the periodogram, \( \langle SR \rangle \) is the mean SR, and $\sigma_{SR}$ is the standard deviation of SR.

In practice, since we evaluate the confidence of the transit signal based on the magnitude of SDE, we often replace \( \textit{SR}_{\text{peak}} \) with \( \textit{SR} \) in Equation~\ref{eq:SDE}. 
This replacement produces the alternative periodogram of \( \textit{SDE} \) versus \textit{trial period}, as shown in Figure~\ref{fig:1_figure}(C).
This transformation makes the detection of the transit signal in the periodogram more intuitive.

Since BLS introduced a breakthrough detection strategy, numerous improvements have emerged. Among them, the TLS method has demonstrated notable gains in detection accuracy.
Specifically, TLS employs a limb-darkened physical transit model, enabling more precise detection of transit signals.
The developers indicate that TLS improves detection efficiency by approximately 10 percent compared to BLS, according to \href{https://github.com/hippke/tls}{TLS GitHub}.

To simulate realistic data observed by the \textit{Kepler} mission, all the light curves we generated contained 65,536 data points, each separated by 29.4244 min. 
The negative dataset was generated by simulating light curves using Gaussian noise, while the positive dataset was constructed by injecting synthetic transit signals according to the pipeline described in Section~\ref{sec:3.2}, as illustrated in Figure~\ref{fig:8figure}.
Ultimately, we generated a customized dataset of light curves with transit SNR \(\in [6, 12]\), ensuring a balanced sample distribution within each transit SNR interval.

In our experiments, the BLS algorithm was implemented using the \textit{autopower} function of the \textit{BoxLeastSquares} class in the \href{https://docs.astropy.org/en/stable/api/astropy.timeseries.BoxLeastSquares.html#astropy.timeseries.BoxLeastSquares.autopower}{\texttt{astropy}} software. 
Specifically, the search period \(\in [100, 150]\) days and the duration \(\in [0.05, 0.93]\) days.
The parameter \textit{oversample} was adjusted to achieve a binned data point count of 4096, and the parameter \textit{frequency factor} was adjusted to achieve a search frequency of 58,400, ensuring consistency with the DELOS settings. 
The TLS algorithm was implemented using the \textit{power} function from the \href{https://transitleastsquares.readthedocs.io/en/latest/Python%20interface.html}{\texttt{transitleastsquares}} software, with the same search period and duration ranges as BLS.
The parameter \textit{oversampling factor} was adjusted to achieve a search frequency of 58,400, maintaining alignment with DELOS.

\subsection{Performance of DELOS versus BLS, TLS}\label{sec:4.2}
We evaluated the performance of DELOS, BLS, and TLS in practical searches by analyzing their ability to distinguish between simulated positive and negative transit signals under different transit SNR conditions.
For BLS and TLS, the maximum SDE in the periodogram was adopted as a quantitative metric to assess the strength of transit signals in the light curves, while DELOS utilized the maximum predicted score in the periodogram. 
By gradually changing the detection threshold over the full range of values for the BLS/TLS SDE and the DELOS score, we evaluated how well each method separates simulated transit signals from no-transit cases under different threshold choices, we obtained a range of discriminative results for simulated positive and negative transit signals.

\begin{figure}
    \centering
    \includegraphics[width=0.8\linewidth]{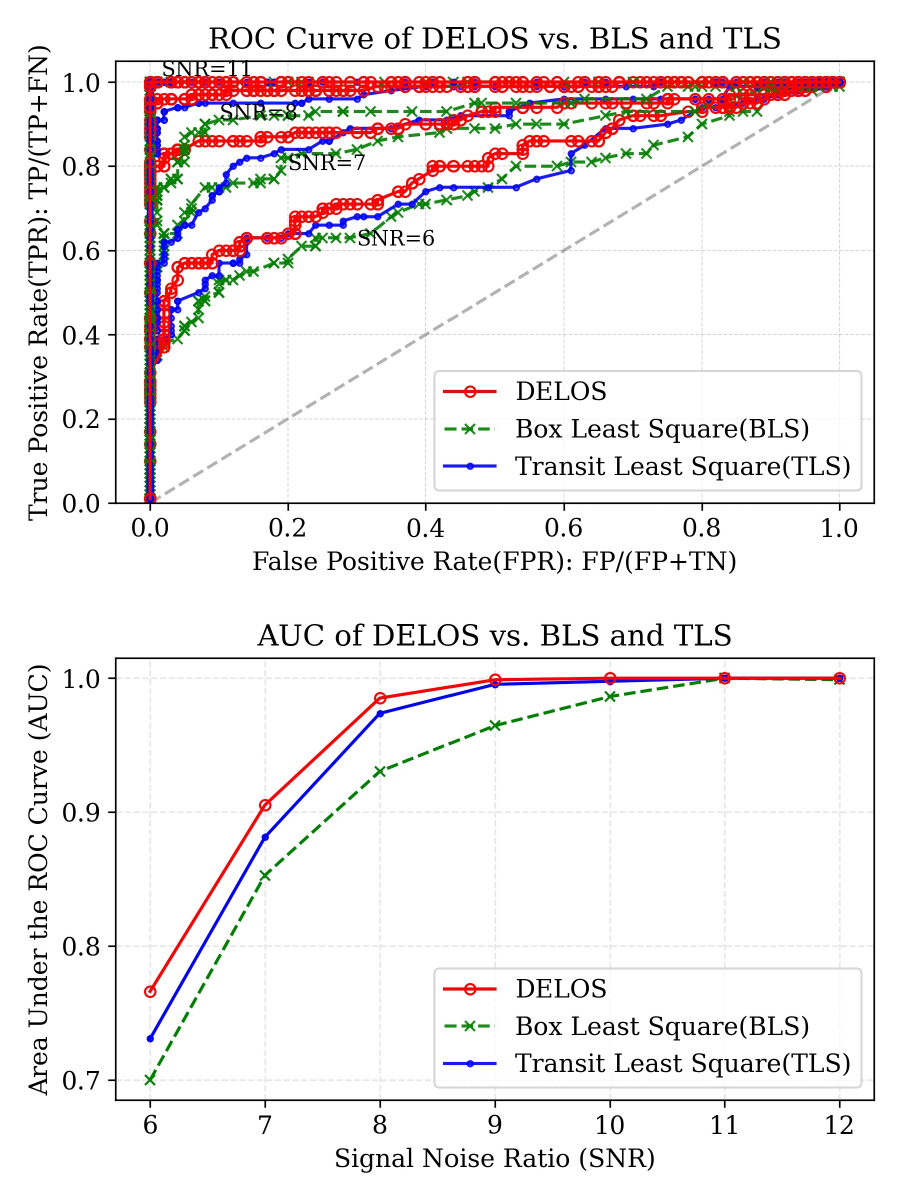}
    \caption{Comparative experimental results of DELOS, BLS, and TLS in terms of Receiver Operating Characteristic (ROC) and Area Under the Curve (AUC). The top panel shows the ROC curves for three methods at transit SNR $\in \{6, 7, 8, 11\}$. The bottom panel shows the AUC of the three methods within the transit SNR $\in [6, 12]$. DELOS achieves higher performance in distinguishing shallow transit from noise.
}
    \label{fig:12figrue}
\end{figure}

\begin{figure}
    \centering
    \includegraphics[width=0.8\linewidth]{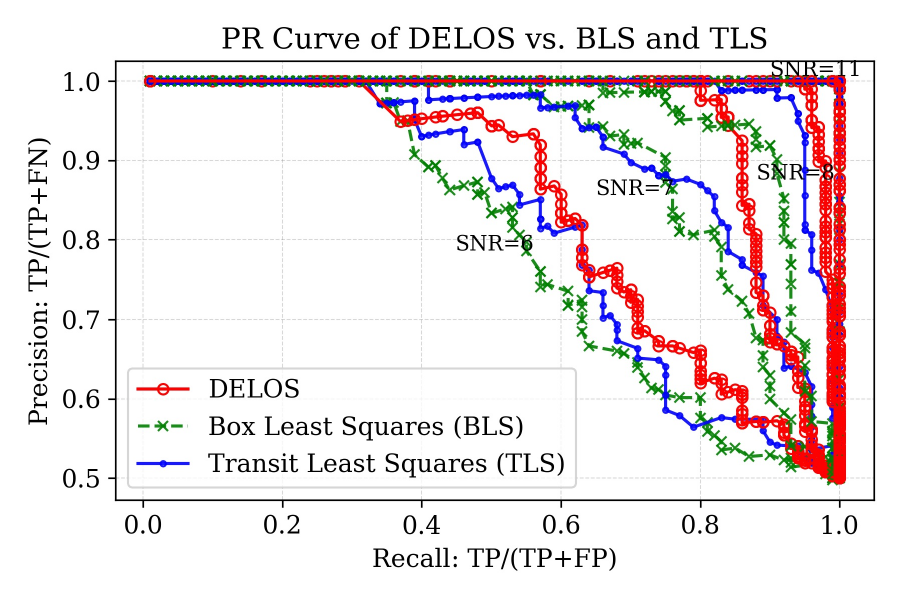}
    \caption{Comparative experimental results of DELOS, BLS, and TLS in terms of PR curve.
    The PR curves for the three methods at transit SNR $\in \{6, 7, 8, 11\}$, DELOS achieves higher precision and recall than BLS and TLS in the low-SNR regime, particularly for transit SNR below 11.
}
    \label{fig:13figrue}
\end{figure}

Based on these results, the false positive rate (FPR) and the true positive rate (TPR) were calculated to construct the ROC curves under different transit SNR, as shown in the upper panel of Figure~\ref{fig:12figrue}.
In the ROC curve, a lower FPR and a higher TPR signify better overall performance.
Furthermore, the AUC serves as a more intuitive metric of the overall capacity to distinguish positive and negative transit signals. 
AUC versus transit SNR for each method is shown in the lower panel of Figure~\ref{fig:12figrue}.
In addition, precision and recall for each discriminative result were calculated, Precision-Recall (PR) curves were constructed, as shown in Figure~\ref{fig:13figrue}. 
In the PR curve, higher recall indicates greater sensitivity to shallow transit signals, while higher precision reflects better reliability. 
Notably, DELOS on average outperforms BLS by 15.5 percent and TLS by 11.25 percent in terms of Precision \& Recall when transit SNR $\in [6,8]$, highlighting its superior ability under low-SNR.

\begin{figure}
    \centering
    \includegraphics[width=0.8\linewidth]{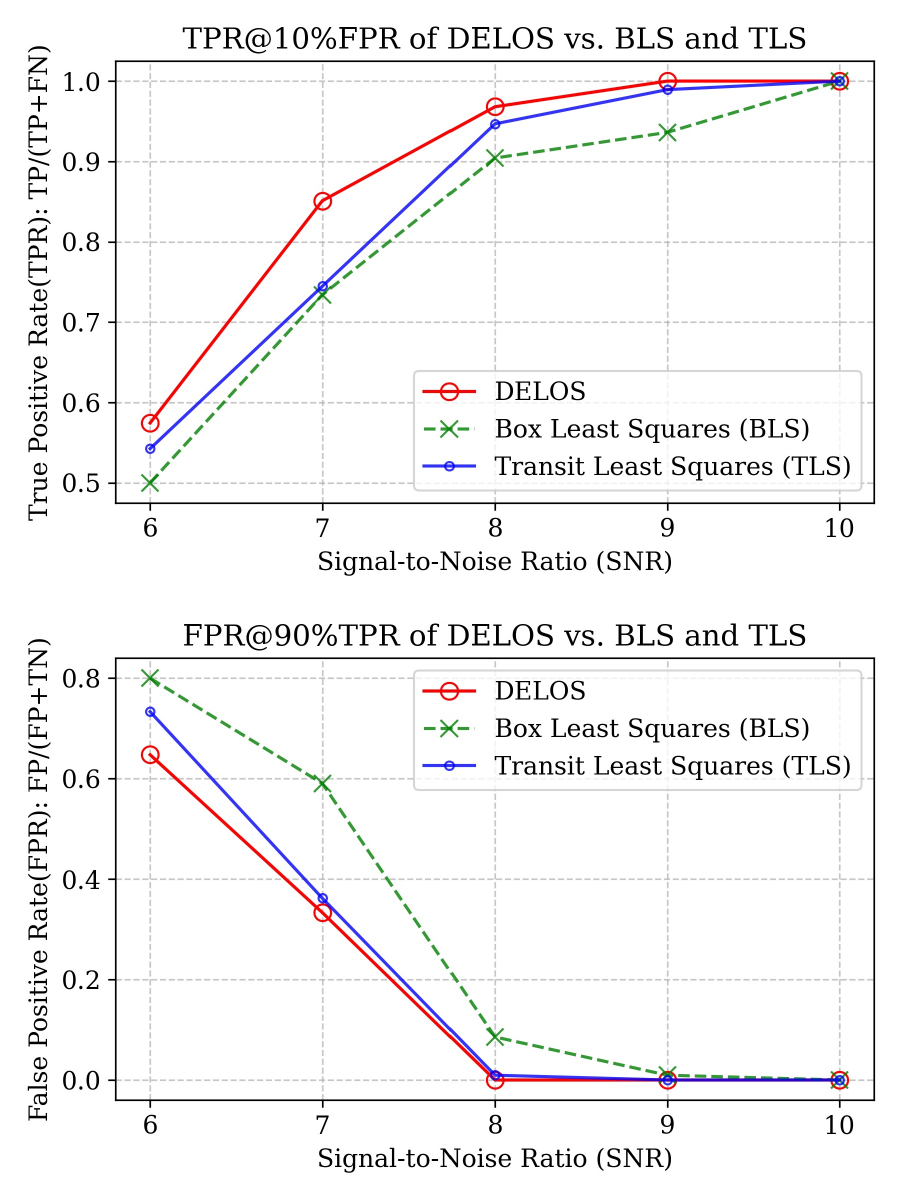}
    \caption{Comparative experimental results of DELOS, BLS, and TLS in terms of the TPR at 10 percent FPR and FPR at 90 percent TPR. 
    DELOS exhibits higher sensitivity and lower false positive rate when transit SNR is below 10.
}
    \label{fig:14figrue}
\end{figure}

However, in practical applications, single-field metrics are often prioritized depending on the specific objective.
For example, when the goal is to detect as many transit signals as possible, methods with higher TPR are preferred because they exhibit greater sensitivity to transit;
conversely, when reducing the false-positive rate is critical, methods with lower FPR are favored because they indicate a lower false-positive rate for transits.
To analyze this trade-off between the three methods, we presented the FPR at 90 percent TPR and the TPR at 10 percent FPR as functions of transit SNR, as shown in Figure~\ref{fig:14figrue}.

Based on the performance comparison among DELOS, BLS, and TLS, we draw the following conclusions.
First, DELOS achieves the best overall performance in distinguishing shallow transit signals from noise.
Second, DELOS shows higher sensitivity to shallow transit signals, as indicated by its higher TPR at a fixed FPR of 10 percent.
Third, DELOS yields fewer false positives, as reflected by its lower FPR at a fixed TPR of 90 percent.
These results suggest that the contrastive-learning-based design of DELOS is well suited not only for shallow transit detection but also for other astronomical applications involving faint signal detection.

Moreover, we have thoroughly analyzed these key reasons behind the superior performance of DELOS, which can be summarized as follows. 
First, DELOS is developed based on contrastive learning, which is highly sensitive to shallow transit signals.
Moreover, this mechanism is also well suited for other scenarios that require the detection of faint signals. 
Second, DELOS employs a realistic transit physical model to simulate transit light curves, thereby allowing the model to learn features closely aligned with observed transit events. 
Third, DELOS was trained on numerous and balanced synthetic data sets, providing robust support for subsequent transit searches.

\subsection{Speed of DELOS versus BLS, TLS}\label{sec:4.3}
\begin{figure}[ht!]
    \centering
    \includegraphics[width=1.0\linewidth]{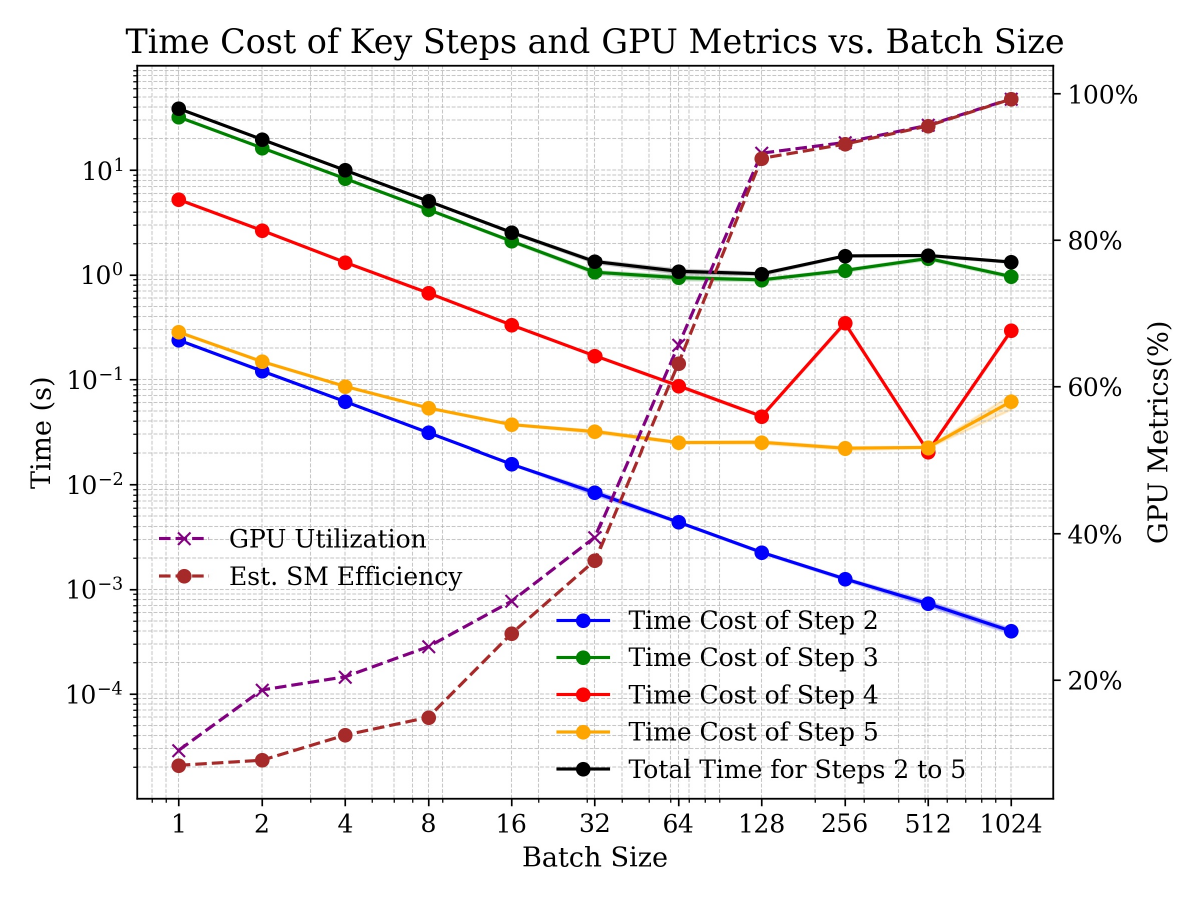}
    \caption{Time cost of key step, \( U_{\text{GPU}} \), and \( \eta_{\text{SM}} \) during the contrastive learning inference stage vary with \textit{BatchSize}.
    The solid lines represent the time cost, using the logarithmic scale on the left axis;
    the dashed lines represent GPU metrics, using the percentage scale on the right axis.
}
    \label{fig:15figrue}
\end{figure}

To evaluate the computational speed of DELOS, we compared it with the BLS algorithm and the TLS algorithm. 
The process by which DELOS produces the score periodogram, together with the GPU profiling metrics used to determine the optimal \textit{BatchSize}, is provided in Appendix~\ref{app:inference_profiling}.
Based on this profiling analysis, we adopt \textit{BatchSize}=128 in all subsequent speed comparisons, for which the contrastive-learning inference stage takes 0.97 seconds, as shown in Figure~\ref{fig:15figrue}.

\begin{figure}
    \centering
    \includegraphics[width=1.0\linewidth]{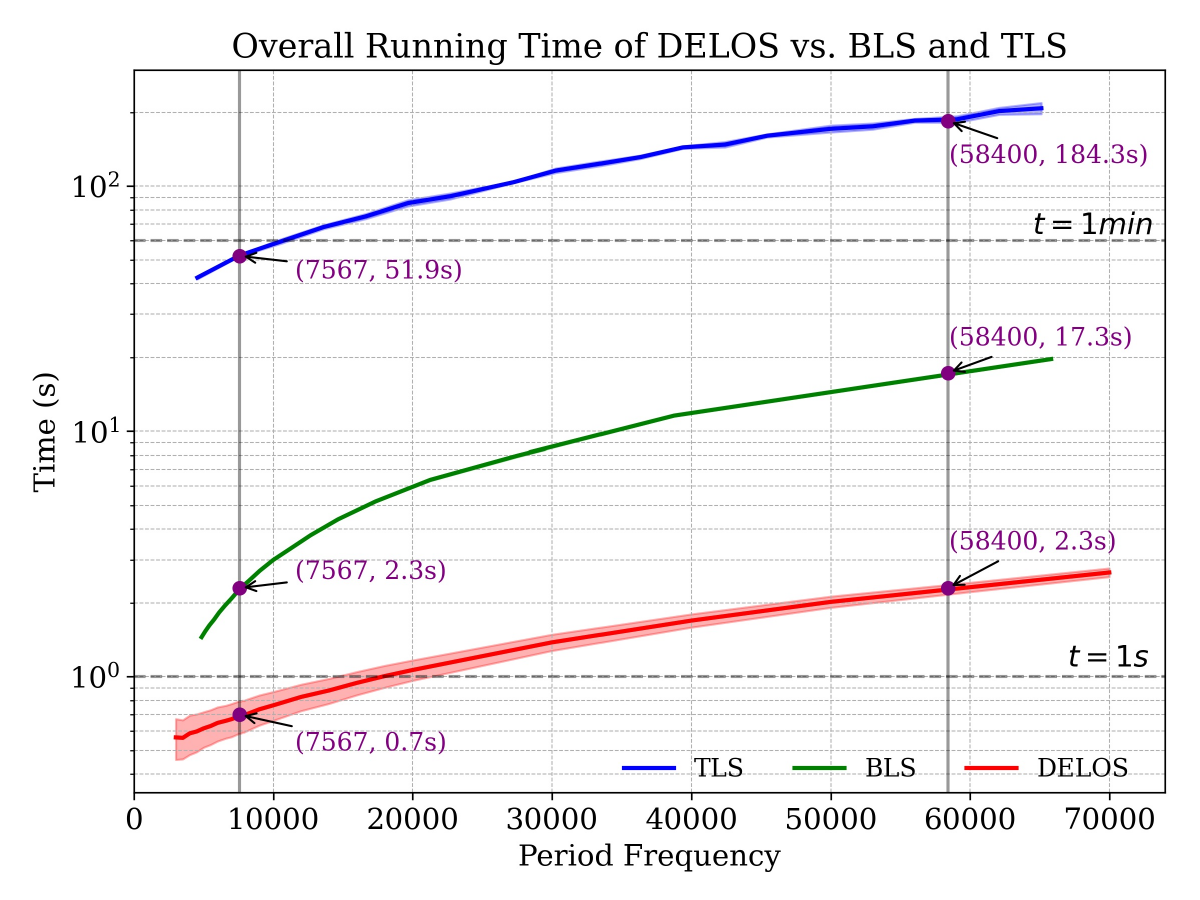}
    \caption{
    Overall runtime of DELOS, BLS, and TLS across different frequencies.
    All three lines are plotted on a logarithmic scale (left axis): red for DELOS, green for BLS, and blue for TLS. The shaded regions indicate the \(\pm 1\sigma\) variation from repeated experiments.
    Additionally, we highlight the overall runtime of the three methods at the ITL sampling resolution of \(N_{\text{sample}} = 58{,}400\), as well as at the cubic sampling frequency proposed in \cite{optical_sampling}, with \(N_{\mathrm{freq,optimal}} = 7{,}567\).}
    \label{fig:16figrue}
\end{figure}

To accelerate BLS computation, we used the fast version implemented in CYTHON.
Subsequently, we evaluated the overall runtime of DELOS, BLS, and TLS at different period frequencies, and the results are presented in Figure~\ref{fig:16figrue}. 
The DELOS runtime is approximately 5 times faster than BLS and around 80 times faster than TLS, and it only takes 2.3 seconds to process a light curve at an ITL period frequency of \( N_{\text{sample}} = 58,400 \).

We also note that \cite{optical_sampling} proposes an optimal frequency sampling strategy for transit signal searches, advocating a cubic (rather than linear) frequency distribution, which can significantly enhance search efficiency. Consequently, to detect ITL low-SNR transits, we compute the optimal frequency sampling distribution using the following formula:

\begin{equation}
\begin{aligned}
f(x) &= \left( \frac{A}{3}x + C \right)^3 \\
C &\equiv f_{\min}^{1/3} - \frac{A}{3} \\
A &\equiv \frac{(2\pi)^{2/3}}{\pi} \cdot \frac{R_\star}{(G M_\star)^{1/3}} \cdot \frac{1}{S \cdot \mathrm{OS}}
\end{aligned}
\label{eq:fx}
\end{equation}
\begin{equation}
N_{\mathrm{freq, optimal}} = \left( f_{\max}^{1/3} - f_{\min}^{1/3} + \frac{A}{3} \right) \cdot \frac{3}{A}
\label{eq:Nfreq}
\end{equation}
Here, \(x\) is the index of the \(x\)-th sampling point. The frequency bounds \(f_{\min}\) and \(f_{\max}\) correspond to \(1/150\) and \(1/100\) day\(^{-1}\), respectively.
\(G\) is the gravitational constant;
\(R_\star\) and \(M_\star\) are the stellar radius and mass, for which we adopt solar values.
\(S\) denotes the total time span of \textit{Kepler} observations (4 years), and \(\mathrm{OS} = 5\) is the oversampling factor. 
Consequently, \(f(x)\) defines the frequency at index \(x\), \(A\) is the frequency growth factor (Hz\(^{1/3}\)/step), and \(N_{\mathrm{freq,optimal}}\) is the total number of sampling points under the cubic sampling scheme.

By substituting all relevant parameters into Equations~\ref{eq:fx} and \ref{eq:Nfreq}, we determined that 7,567 sampling points are required under the cubic distribution.
In Figure~\ref{fig:16figrue}, we also compare the runtime of three methods applied to the same light curve using this sampling scheme. 
Notably, DELOS breaks the 1-second runtime barrier, completing the search in just 0.7 seconds;
in contrast, BLS and TLS take 2.3 seconds and 51.9 seconds, respectively, indicating that DELOS is approximately 3 times and 74 times faster than BLS and TLS under the same sampling conditions.

Furthermore, when using DELOS to search for planets around approximately 156,000 main-sequence stars observed by the \textit{Kepler} mission, the estimated processing time is about 1.26 to 4.15 days.
Essentially, the reasons for the faster speed of DELOS can be attributed to two main factors. 
First, DELOS has fewer loops and simpler computational rules in its algorithm structure. 
Second, DELOS enables parallel processing on GPU machines by setting an appropriate \textit{BatchSize}, significantly reducing overall runtime without sacrificing accuracy.
In the future, with the continuous upgrade of GPU hardware, the evaluation process will become even faster.

\section{DELOS applied to real \textit{Kepler} data}\label{sec:5}
\subsection{\textit{Kepler} DR25 Light Curve Processing}\label{sec:5.1}
To test the performance of DELOS on real \textit{Kepler} data, we downloaded light curves from Q1-Q17 \textit{Kepler} Data Release 25 (DR25), available via the \textit{Kepler} archive at the Mikulski Archive for Space Telescopes (\href{https://archive.stsci.edu/}{\texttt{MAST}}).
These light curves were then processed using the data preprocessing pipeline in Section~\ref{sec:3.1}.
The primary focus herein is on the detrending preprocessing step for already cotrended light curves.
While no unified standard has been established for this step when processing \textit{Kepler} data to date, the pipeline presented herein was developed in-house, following the processing workflow outlined in \cite{gpfc_1}.
The result of this approach is demonstrated in Figure~\ref{fig:2_figure}.

\subsection{Construction of Real \textit{Kepler} Validation Samples}\label{sec:5.2}
We downloaded the KOI archive dated January 21, 2025, which contains 9,564 targets, each with available raw PDCSAP long-cadence light curves.
This archive comprises 2,743 confirmed planets, 1,982 candidate planets, and 4,839 false positives. 
We also downloaded the corresponding transit parameters, including orbital period, transit midpoint, and transit duration;
these were used for extensive testing and validation.

Subsequently, we selected ITL low-SNR transit signals with exoplanetary archive disposition as ``CONFIRMED'' to construct the positive data set while ensuring that each planetary system contained only a single star.
For each light curve in the positive data set, we retained the real ITL transit signal and removed other periodic signals reported in the KOI archive. 
Meanwhile, we selected ``CONFIRMED'' or ``CANDIDATE'' transit signals and removed all periodic signals recorded in KOI to obtain no-transit light curves, adhering to the same single-star system restriction.
After removing duplicate stellar light curves, we finally obtained 3,493 no-transit light curves.
Due to the lack of details on ``FALSE POSITIVE'' in KOI, we excluded planetary systems with ``FALSE POSITIVE'' signals.

Finally, we obtained a positive data set consisting of 60 light curves, which were used to evaluate the recovery rate of DELOS for real \textit{Kepler} ITL transit signals.
Accordingly, we also obtained a negative data set consisting of 3,493 light curves, which were used for transit searches by DELOS.

\subsection{Recovery of Confirmed \textit{Kepler} ITL Exoplanets}\label{sec:5.3}
With light curves and transit parameters from the KOI archive, we applied DELOS to the positive dataset to recover known transit signals.
The results are summarized in Table~\ref{table:1}, where for each target we list the period, depth, and SNR reported by KOI, together with the period and score identified by DELOS. 

\begin{table*}[t]
\centering
\scriptsize
\caption{Summary of all known ITL, low-SNR transit signals recovered by DELOS.}
\label{table:1}
\setlength{\tabcolsep}{4pt}
\begin{tabular}{lccccccc}
\toprule
Kepler Name & Transit Period & Transit SNR & Transit Depth & TLS SDE & DELOS Period & Folded Idx & DELOS Score \\
\midrule
1 Kepler-1801 c & 116.58 & 8.7 & 819 & 5.7 & 116.58 & 19370 & 0.91 \\
2 Kepler-1620 b & 101.95 & 10.3 & 153 & 13.0 & 101.95 & 02282 & 1.00 \\
3 Kepler-1653 b & 140.25 & 11.9 & 568 & 17.3 & 140.25 & 47013 & 1.00 \\
4 Kepler-1599 b & 122.36 & 12.2 & 353 & 10.5 & 122.36 & 26114 & 0.95 \\
5 Kepler-1609 b & 114.34 & 13.0 & 240 & 8.1 & 114.34 & 16750 & 0.92 \\
6 Kepler-186 f & 129.95 & 13.1 & 500 & 21.8 & 129.95 & 34983 & 0.95 \\
7 Kepler-442 b & 112.30 & 13.1 & 502 & 22.9 & 112.30 & 14370 & 1.00 \\
8 Kepler-351 d & 142.54 & 13.7 & 905 & 18.0 & 142.54 & 49692 & 1.00 \\
9 Kepler-430 c & 110.98 & 14.0 & 150 & 9.3 & 110.98 & 12830 & 1.00 \\
10 Kepler-722 c & 105.15 & 15.4 & 685 & 23.4 & 105.15 & 06010 & 1.00 \\
11 Kepler-1185 b & 104.35 & 15.8 & 244 & 6.7 & 104.32 & 05040 & 0.91 \\
12 Kepler-132 e & 110.29 & 16.4 & 145 & 17.2 & 110.29 & 12010 & 1.00 \\
13 Kepler-1548 b & 124.83 & 17.1 & 648 & 29.8 & 124.83 & 28998 & 1.00 \\
14 Kepler-1883 b & 103.43 & 17.8 & 950 & 31.8 & 103.43 & 04003 & 0.98 \\
15 Kepler-1362 b & 136.21 & 20.0 & 1209 & 34.7 & 136.21 & 42287 & 1.00 \\
\midrule
16 Kepler-1539 b & 133.30 & 20.6 & 1018 & 29.2 & 133.30 & 38898 & 1.00 \\
17 Kepler-1294 b & 115.69 & 20.7 & 946 & 25.9 & 115.69 & 18321 & 1.00 \\
18 Kepler-281 d & 148.27 & 21.8 & 1707 & 41.4 & 148.27 & 56381 & 1.00 \\
19 Kepler-549 c & 117.03 & 21.9 & 770 & 30.0 & 117.03 & 19895 & 1.00 \\
20 Kepler-1540 b & 125.41 & 22.6 & 1345 & 24.2 & 125.41 & 29680 & 0.92 \\
21 Kepler-309 c & 105.36 & 22.6 & 1210 & 38.7 & 105.36 & 06258 & 1.00 \\
22 Kepler-196 d & 122.08 & 23.3 & 800 & 39.9 & 122.08 & 25790 & 1.00 \\
23 Kepler-344 c & 125.60 & 23.3 & 926 & 34.0 & 125.60 & 29903 & 1.00 \\
24 Kepler-1840 b & 131.19 & 23.9 & 916 & 30.0 & 131.19 & 36427 & 1.00 \\
25 Kepler-1177 b & 106.25 & 24.0 & 397 & 24.2 & 106.25 & 07297 & 1.00 \\
26 Kepler-440 b & 101.11 & 24.1 & 878 & 31.5 & 101.11 & 01297 & 1.00 \\
27 Kepler-1126 b & 108.59 & 24.3 & 338 & 31.3 & 108.59 & 10037 & 1.00 \\
28 Kepler-90 f & 124.92 & 24.4 & 499 & 7.4 & 124.92 & 29108 & 1.00 \\
29 Kepler-1936 b & 126.56 & 24.8 & 366 & 31.0 & 126.56 & 31018 & 0.91 \\
30 Kepler-1810 b & 101.13 & 25.7 & 1503 & 39.6 & 101.13 & 01323 & 1.00 \\
\midrule
31 Kepler-1678 b & 147.97 & 26.4 & 1267 & 35.1 & 147.97 & 56033 & 1.00 \\
32 Kepler-1333 b & 109.65 & 26.7 & 900 & 38.0 & 109.65 & 11267 & 1.00 \\
33 Kepler-896 b & 144.55 & 27.1 & 1035 & 21.0 & 144.55 & 52035 & 1.00 \\
34 Kepler-1535 b & 138.94 & 28.0 & 530 & 29.2 & 138.94 & 45486 & 1.00 \\
35 Kepler-264 c & 140.11 & 29.0 & 267 & 23.3 & 140.11 & 46845 & 1.00 \\
36 Kepler-1341 b & 132.99 & 29.3 & 1275 & 35.3 & 132.99 & 38532 & 1.00 \\
37 Kepler-1780 b & 109.45 & 30.1 & 1270 & 43.6 & 109.45 & 11032 & 1.00 \\
38 Kepler-266 c & 107.72 & 30.5 & 1396 & 29.4 & 107.72 & 09018 & 1.00 \\
39 Kepler-985 b & 116.33 & 31.5 & 1586 & 31.2 & 116.33 & 19074 & 0.95 \\
40 Kepler-1038 b & 148.46 & 33.7 & 1532 & 39.4 & 148.46 & 56601 & 1.00 \\
41 Kepler-1766 b & 105.01 & 33.9 & 296 & 34.6 & 105.01 & 05857 & 1.00 \\
42 Kepler-920 c & 100.83 & 36.7 & 1779 & 28.9 & 100.83 & 00967 & 1.00 \\
43 Kepler-841 b & 124.42 & 39.9 & 4107 & 48.4 & 124.42 & 28522 & 1.00 \\
44 Kepler-987 b & 105.30 & 40.6 & 1362 & 39.3 & 105.30 & 06194 & 1.00 \\
45 Kepler-1036 b & 122.88 & 41.9 & 718 & 33.7 & 122.88 & 26724 & 1.00 \\
\midrule
46 Kepler-1058 b & 110.96 & 42.1 & 1283 & 29.7 & 110.96 & 12806 & 1.00 \\
47 Kepler-1016 c & 105.65 & 42.3 & 1367 & 38.1 & 105.65 & 06605 & 1.00 \\
48 Kepler-1000 b & 120.02 & 46.1 & 886 & 28.4 & 120.02 & 23368 & 0.92 \\
49 Kepler-62 e & 122.39 & 46.3 & 711 & 38.3 & 122.39 & 26146 & 1.00 \\
50 Kepler-218 d & 124.52 & 47.5 & 690 & 33.6 & 124.52 & 28644 & 1.00 \\
51 Kepler-976 b & 105.96 & 51.8 & 2286 & 35.3 & 105.96 & 06963 & 1.00 \\
52 Kepler-965 b & 134.25 & 57.2 & 832 & 31.7 & 134.25 & 40012 & 1.00 \\
53 Kepler-397 c & 135.49 & 73.2 & 5038 & 42.8 & 135.49 & 41461 & 1.00 \\
54 Kepler-820 b & 127.83 & 76.9 & 1830 & 28.7 & 127.83 & 32509 & 0.99 \\
55 Kepler-126 d & 100.28 & 81.9 & 351 & 29.7 & 100.28 & 00331 & 1.00 \\
56 Kepler-11 g & 118.38 & 83.1 & 1083 & 32.0 & 118.38 & 21466 & 1.00 \\
57 Kepler-1718 b & 135.19 & 85.8 & 663 & 34.9 & 135.19 & 41099 & 1.00 \\
58 Kepler-952 b & 130.35 & 103.7 & 5129 & 40.7 & 130.35 & 35453 & 1.00 \\
59 Kepler-302 c & 127.28 & 148.1 & 8089 & 37.2 & 127.28 & 31865 & 1.00 \\
60 Kepler-1662 b & 134.48 & 149.9 & 4034 & 42.9 & 134.48 & 40270 & 0.93 \\
\bottomrule
\end{tabular}

\vspace{2pt}
\begin{minipage}{\textwidth}
\scriptsize
\textit{Note.}
Targets are selected from the KOI catalog \citep{robotvetter} with $P\in[100,150]$\,d, Transit SNR$<150$, Transit Depth$<10{,}000$\,ppm, and labeled as confirmed planets.
TLS SDE is computed using the TLS algorithm \citep{TLS}.
The DELOS period is derived from the folded index $i$ via $P=\frac{50}{58{,}400}\,i+100$\,d.
\end{minipage}
\end{table*}

\begin{figure}[ht!]
    \centering
    \includegraphics[width=1.0\linewidth]{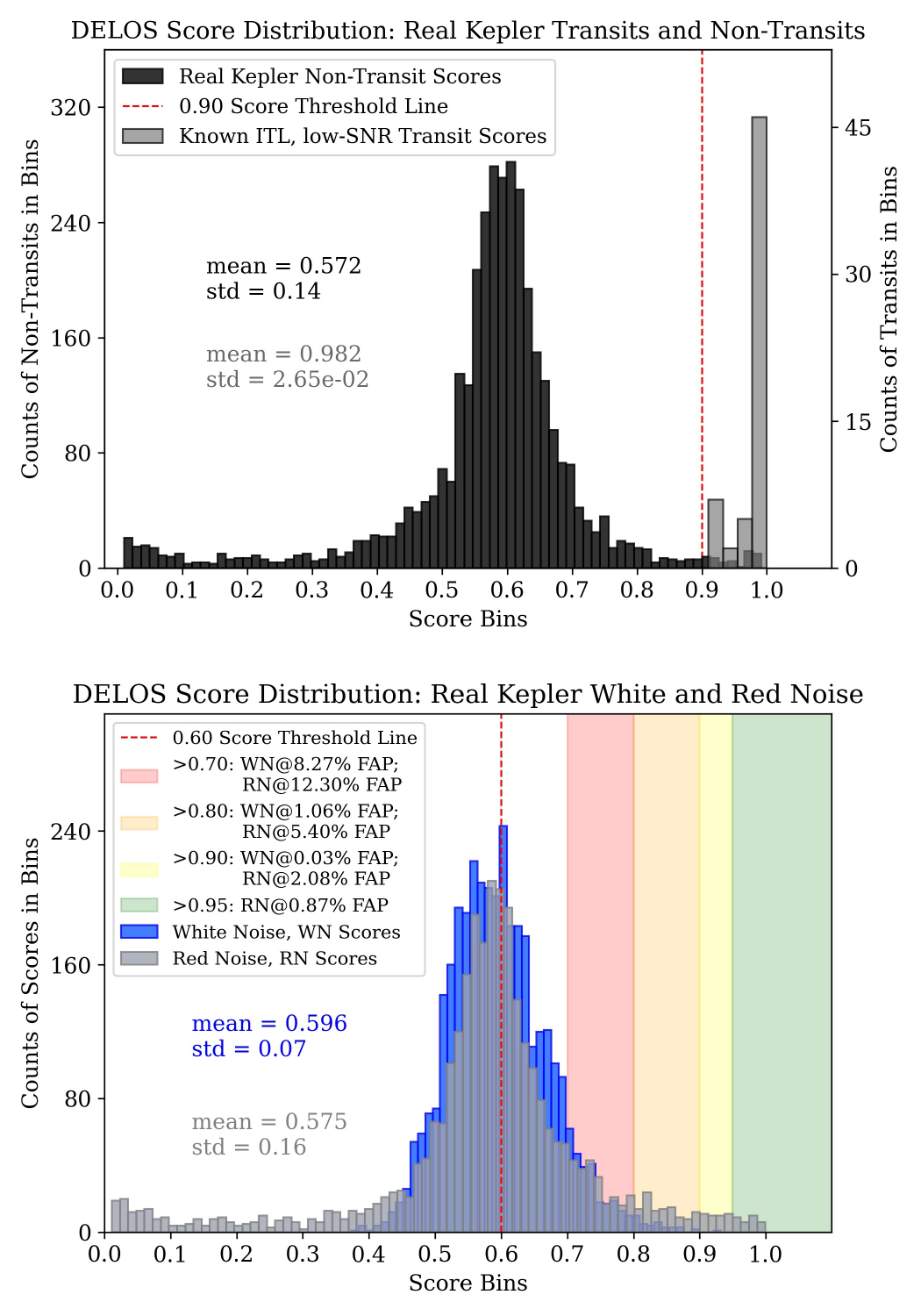}
    \caption{DELOS score distributions for real \textit{Kepler} samples and noise-control tests. The upper panel compares the scores of 3493 KOI non-transit samples with those of 60 known \textit{Kepler} ITL, low-SNR transits, selected as described in Section~\ref{sec:5.2}. The red dashed line marks the adopted score threshold of 0.90. This threshold also recovers all 60 known ITL, low-SNR transit signals. The lower panel shows the score distributions for Kepler-like white-noise and red-noise control samples. The red dashed line marks the 60\% score threshold line. The white-noise samples are generated by simulating the empirical \textit{Kepler} noise distribution under a Gaussian assumption, whereas the red-noise samples are constructed by removing known transit signals from the 3493 real samples and applying time-axis shuffling to the residual light curves. The shaded regions indicate score intervals associated with different False-Alarm Probability (FAP) levels. For the white-noise controls, scores above 0.80 correspond to an FAP of approximately 1\%, while for the red-noise controls, scores above 0.95 are required to reduce the FAP below 1\%. These results indicate that the DELOS score can serve as an effective ranking statistic for separating transit-like signals from noise-induced false alarms in real \textit{Kepler} data.}
    \label{fig:18figrue}
\end{figure}

Importantly, DELOS recovers all 60 confirmed ITL low-SNR signals in the correct period with scores $>0.90$.
The upper panel of Figure~\ref{fig:18figrue} shows the score distribution ($mean \sim 0.99$; $\sigma \sim 0.03$); 
the recovered periods deviate from the KOI values by $<0.01\%$.
These results verify all previously known ITL low-SNR exoplanet signals, demonstrate the utility of DELOS for practical surveys, and establish a solid foundation for discovering unknown ITL signals.

\begin{figure}[ht!]
    \centering
    \includegraphics[width=0.9\linewidth]{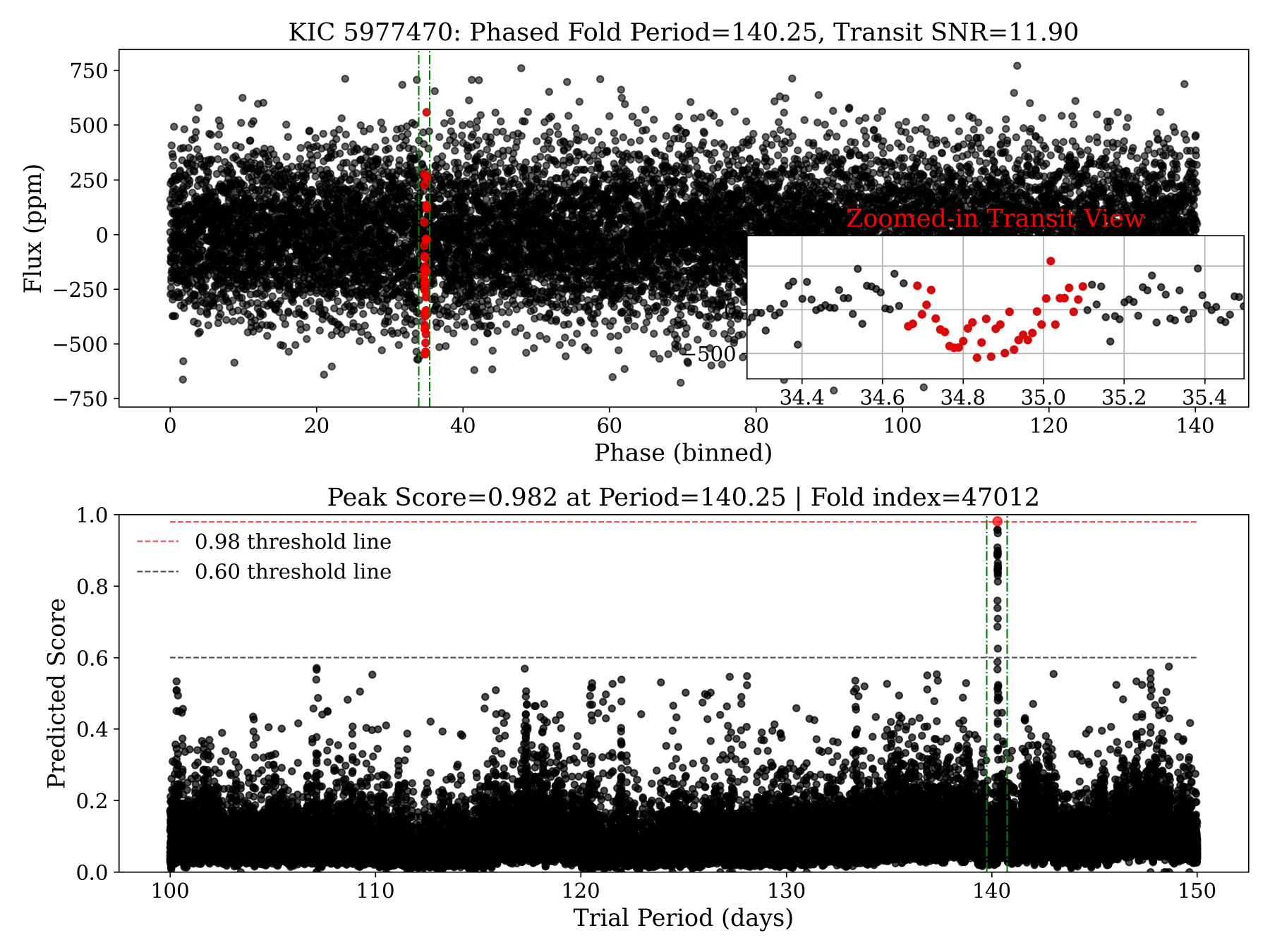}
    \caption{Example of a \textit{Kepler} ITL exoplanetary signal recovered by DELOS. 
    The upper panel presents the folded light curve of Kepler-1653 b at the correct transit period of 140.25 days, and the inset in the lower right corner provides a zoomed-in view of this transit.
    The lower panel displays the score periodogram produced by DELOS, revealing a prominent peak at 140.25 days with 1.00 score, precisely aligning with the TRUE period.
}
    \label{fig:17figrue}
\end{figure}

To further illustrate the ability of DELOS to identify real transits, KIC 5977470 (Kepler-1653 b) is selected as an example.
This system has a transit signal with a period of 140.25 days and a transit SNR of 11.90.
The folded light curve and the DELOS score periodogram are illustrated in Figure~\ref{fig:17figrue}. 
The maximum score is 1.00, corresponding to the 47,012th index among 58,400 trial periods.
This result indicates that DELOS estimates the transit period as \(\frac{47,012}{58,400} \times 50 + 100 = 140.25\) days. 
This estimate precisely aligns with the TRUE period recorded in the KOI archive.

\subsection{Search for Additional High-scoring ITL Transit-like Signals}\label{sec:5.4}
Subsequently, DELOS was deployed on the negative dataset, from which all reported periodic signals in KOI were removed.
For each light curve, we selected the maximum score from the searched score periodogram, with the additional requirements that, to suppress spuriously high scores caused by accidental noise folding, the peak exceeds the \(3\sigma\) upper limit of the remaining scores and has a width of at least five data points.

We used DELOS to search 3,493 stellar targets (see Section~\ref{sec:5.2}), yielding 3,493 scores.
Their distribution is shown by the black histogram in the upper panel of Figure~\ref{fig:18figrue}.
We found that the majority of no-transit light curves produced scores below the 0.70 threshold. 
In contrast, DELOS identified periodic signals exceeding this threshold as potential transit candidates.

Among these high-scoring signals, further scrutiny will be conducted, potentially revealing some as genuine ITL transit candidates. 
However, they require independent recovery, contamination analysis, and full astrophysical vetting before any planetary interpretation can be claimed.

\subsection{Robustness of DELOS to Realistic White Noise}
\label{sec:5.5}
To quantify the FAP of DELOS for high-scoring detections, we first constructed a white-noise control sample designed to mimic the observational baseline and noise amplitude of real \textit{Kepler} light curves. 
After the preprocessing procedure described in Section~\ref{sec:3.1}, we assume in this test that the residual light curves are dominated by approximately Gaussian white noise. 
Each simulated light curve contains 65,536 data points with a cadence of 29.4244 minutes, corresponding to the four-year observing window of the \textit{Kepler} mission. 
The noise amplitude of each curve was randomly sampled from the empirical noise distribution of real ITL light curves shown in Figure~\ref{fig:7figure}. 
In total, we generated 3,493 white-noise light curves, matching the number of non-transit samples used in Section~\ref{sec:5.2}, and processed them with the same DELOS search pipeline described in Section~\ref{sec:5.3}.

The resulting score distribution is shown in the lower panel of Figure~\ref{fig:18figrue}. 
Using score thresholds of $>0.70$, $>0.80$, and $>0.90$, DELOS identifies 289, 37, and 1 white-noise light curves as transit-like signals, respectively. 
The corresponding empirical FAPs are $289/3493 = 8.27\%$, $37/3493 = 1.06\%$, and $1/3493 = 0.03\%$. 
Under the white-noise assumption, a score threshold of 0.90 is sufficient to suppress the false-alarm probability to well below 1\%. 
This result indicates that high DELOS scores are unlikely to be produced by purely white-noise fluctuations in the adopted validation setting.

\subsection{Robustness of DELOS to Realistic Red Noise}
\label{sec:5.6}
Although the white-noise experiment demonstrates a low FAP at a score threshold of 0.90, this assumption is likely optimistic for real \textit{Kepler} observations. 
The actual \textit{Kepler} background can contain correlated noise, residual stellar variability, quarter-boundary effects, instrumental systematics, harmonics, aliases, and contamination from nearby sources. 
Therefore, we further constructed a red-noise control sample to evaluate whether DELOS can distinguish genuine transit-like signals from more realistic noise-induced periodic patterns.

For this test, we started from the 3,493 real \textit{Kepler} samples used in Section~\ref{sec:5.2}. 
Known transit signals were first removed from these light curves. 
We then applied a target-wise time-shuffling procedure to the residual light curves, preserving the empirical residual-noise distribution of each target while destroying coherent transit periodicity. 
The resulting red-noise control sample was processed using the same DELOS search pipeline as in the white-noise experiment.

The red-noise score distribution is also shown in the lower panel of Figure~\ref{fig:18figrue}. 
At score thresholds of $>0.70$, $>0.80$, $>0.90$, and $>0.95$, DELOS identifies 429, 189, 73, and 30 red-noise light curves as transit-like signals, respectively. 
The corresponding empirical FAPs are $429/3493 = 12.30\%$, $189/3493 = 5.40\%$, $73/3493 = 2.08\%$, and $30/3493 = 0.87\%$. 
Compared with the white noise case, the red-noise control sample produces a broader high-score tail, confirming that realistic residual structures are more challenging than Gaussian white noise. 
Nevertheless, increasing the score threshold to 0.95 reduces the red-noise FAP below 1\%, demonstrating that DELOS retains strong discriminatory power against noise-induced false alarms in realistic \textit{Kepler} residuals.

In practical searches, a lower score threshold may still be adopted to avoid missing weak long-period transit candidates, followed by additional vetting and validation procedures. 
The red-noise experiment, therefore, provides a conservative noise-control test rather than a final planet-validation criterion and supports the use of the DELOS score as a ranking statistic for prioritizing low-SNR ITL transit candidates.

\section{Discussion}\label{sec:6}
\subsection{Improvements from Cotrending and Detrending}\label{sec:6.1}
With the advent of the \textit{Kepler} mission, the transit method has become the most prolific technique for exoplanet detection. 
However, despite extensive observations over the years, a bona fide Earth twin remains elusive, primarily due to the inherently long periods, shallow, and narrow features of Earth-like transit signals. 
To overcome this challenge, beyond the design of high-sensitivity detection algorithms to enhance the identification of shallow signals, enhancing the noise suppression capability of the Cotrending and Detrending algorithms within the \textit{Kepler} pipeline is another crucial strategy.  
These preprocessing algorithms critically shape the initial light curve fidelity on which all subsequent detection methods are based. 
Therefore, integrating more sophisticated noise suppression techniques (e.g. \citealt{Cotrending_imp1, Detrending_imp1}) into the Cotrending and Detrending processes would markedly elevate the quality of \textit{Kepler} data while substantially increasing the probability of detecting Earth-sized exoplanets in subsequent analyzes.

\subsection{Further Speed Tests}\label{sec:6.2}
We noticed the development of faster transit detection algorithms, such as GPU-accelerated BLS \href{https://johnh2o2.github.io/cuvarbase/index.html}{GBLS} and GTLS (Q. Hu et al. 2026, in preparation), which employed GPU parallel computing to accelerate internal computations without compromising detection performance. 
To assess the computational speed of DELOS relative to these algorithms, we conducted speed tests using the same dataset, computing hardware, and parameter configurations as described in Section~\ref{sec:4.2}.
The experimental results show that GBLS and GTLS required only 0.7~s and 18.3~s, respectively, to process a single light curve at a sampling frequency of \(N_{\text{sample}} = 58{,}400\).
However, despite its higher speed, GBLS achieved worse overall performance than DELOS.
In addition, we noticed that while the runtime of each algorithm was influenced to some extent by the hardware configuration, the relative speed differences among the algorithms remained consistent.

\section{Conclusion and future directions}\label{sec:7}
We have presented \textbf{DELOS}, a contrastive-learning approach for identifying intermediate-to-long-period shallow transit signals in \textit{Kepler} photometry.
Unlike many neural-network vetters that classify pre-detected threshold-crossing events, DELOS directly scores phase-folded light curves over a dense grid of trial periods and produces a score periodogram.
It can therefore operate as a first-stage transit-search engine rather than only as a post-detection classifier.

For the initial 100–150 day search window, DELOS was trained on 20 million synthetic light curves generated with realistic transit models and Kepler-like noise properties.
The trained model achieved a validation accuracy of 99.3\% on synthetic low-SNR light curves.
In controlled injection-recovery experiments, DELOS improved the combined precision–recall performance by 15.5\% relative to BLS and 11.25\% relative to TLS.
It also provided substantial computational speedups, running approximately 3–5 times faster than BLS and 74–80 times faster than TLS in the tested configuration.
Applied to a selected \textit{Kepler} validation sample, DELOS recovered all 60 known shallow intermediate-to-long-period signals in the targeted period range.

These results demonstrate that contrastive representation learning, combined with GPU-accelerated phase folding and optimized phase binning, provides an efficient and sensitive approach for searching for low-SNR transit signals.
At the same time, DELOS scores should be interpreted as transit-likeness rankings rather than calibrated astrophysical probabilities.
High-scoring signals identified by DELOS require independent recovery with other search algorithms, contamination checks, temporal-consistency tests, false-alarm analysis, and full astrophysical vetting before they can be considered validated planet candidates.

The present work should therefore be viewed as a methodological development and validation study.
Future improvements will focus on extending the framework to longer orbital periods, improving robustness to correlated noise and stellar variability, optimizing detrending and cotrending procedures, and applying the method to larger samples from \textit{Kepler}, \textit{K2}, and \textit{TESS}, as well as future missions such as \textit{PLATO} \citep{plato1, plato2} and \textit{ET} \citep{et1, et2, et2_2024}.
With these extensions, DELOS may become a useful tool for expanding the search for shallow, terrestrial-sized planets in long-baseline photometric surveys.

\begin{acknowledgements}

Funding for this study is provided by the Strategic Priority Program on Space Science of the Chinese Academy of Sciences (XDA15020600) and China's Space Origins Exploration Program (GJ11030405).
XYY acknowledges the support of the Young Scientists Fund of the National Natural Science Foundation of China, Grant No. 12303068, and the Special Research Assistant Program of the Chinese Academy of Sciences for high-precision photometric research with the Earth 2.0 satellite, Grant No. E283071009; JPZ is also grateful for the support of the National Natural Science Foundation of China, Grant No. 12203087.
We thank Bo Ma, Shiyin Shen, Luoxi Jin, Wanchang Xiao, and Hao Ding for valuable suggestions that improved the manuscript.

The public photometric data analyzed here are the Q1--Q17 long-cadence PDCSAP light curves from the \textit{Kepler} mission archive at the Mikulski Archive for Space Telescopes (MAST).
The KOI and exoplanet parameters used in this work were obtained from the NASA Exoplanet Archive, including the KOI cumulative table \citep{NEA_KOI_Cumulative}, accessed on 2025 January 21.
The derived light curves, simulated light curves, trained DELOS model, preprocessing and simulation scripts, score-periodogram outputs, benchmark outputs, and figure-source data are based on these public data and will be made available upon reasonable request from the corresponding author.

\end{acknowledgements}

\bibliographystyle{aa}
\bibliography{literature}

\begin{appendix}




\twocolumn

\section{Contrastive learning in DELOS}
\label{app:contrastive_components}
As described in Section~\ref{sec:2.2}, DELOS follows the general design of contrastive learning frameworks and consists of the data augmentation module, the encoder network, the projection network, and the contrastive loss. 
This appendix provides the detailed roles of these components in the DELOS framework, while the main text focuses on the motivation of contrastive learning and the supervised contrastive objective.

\begin{itemize}
\setlength{\itemsep}{2pt}

    \item 
    A stochastic \textit{data augmentation} module, referred to as \textit{Aug($\cdot$)} in Figure~\ref{fig:3figure}, generates different augmented views from the same initial transit or noise sample \( x \).
    These augmented views form a transit or noise pair, and in the DELOS framework, each pair consists of up to 100 augmented views. 
    For simplicity, we illustrate this module using two views in a pair, denoted as \( \tilde{x}_1 \), \( \tilde{x}_2 \), both generated from \( x \), where \( \tilde{x}_1 = \text{\textit{Aug}}(x) \), \( \tilde{x}_2 = \text{\textit{Aug}}(x) \). 
    Each view captures a distinct perspective of the underlying shallow transit signal or noise realization while retaining partial original information. 
    In the DELOS framework, generating more diverse augmented views from the same sample facilitates robustness against observational noise and systematic variations \citep{CLS}.

    \item A flexible \textit{encoder network}, referred to as \( \text{\textit{Enc}}(\cdot) \) in Figure~\ref{fig:3figure}, extracts representations from augmented views \( \tilde{x} \). 
    This encoder outputs high-quality, compact distributional representations, which help DELOS enhance the separability between shallow transit signals and noise.
    To illustrate this process, we denote the encoded representations as \( \boldsymbol{h}_1 = \text{\textit{Enc}}(\tilde{x}_1) \) and \( \boldsymbol{h}_2 = \text{\textit{Enc}}(\tilde{x}_2) \), where \( \boldsymbol{h}_{1,2} \in \mathbb{R}^{D_e} \), \(\mathbb{R}^{D_e} \) is the \( D_e \)-dimensional REAL vector.
    In the DELOS framework, the best performance was achieved with \( D_e = 2048 \).
    The encoder network is typically initialized from a pre-trained Transformer-based model that has been trained to convergence on large-scale ImageNet, as such pre-training enables the encoder to learn compact and transferable distributional representations \citep{ResNet}.
    However, transit detection is fundamentally an anomaly detection problem in one-dimensional time-series data, for which no suitable large-scale pre-trained encoder is currently available.
    Therefore, DELOS employs a custom-designed CNN-based encoder that is trained entirely on our simulated dataset and tailored to capture the characteristic structures of shallow transit signals under realistic noise conditions.

    \item 
    An auxiliary but indispensable \textit{projection network}, referred to as \( \textit{Proj}(\cdot) \) in Figure~\ref{fig:3figure}.
    DELOS includes two losses: the contrastive loss and the classification loss. They are jointly computed to update the network parameters via backpropagation. 
    The contrastive loss optimizes the encoder and projection networks, while the classification loss is used only to optimize the classifier network, keeping the encoder and projection parameters fixed.
    These two objectives are not necessarily aligned in their optimization directions and may introduce objective interference when optimized on the same representation space \citep{CLS}.
    Therefore, the encoder output representation is used to compute the classification loss, while we project the encoder output representations \( \boldsymbol{h} \) into more compact representations \( z \) to compute the contrastive loss, i.e., \( z_1 = \textit{Proj}(\boldsymbol{h}_1) \) and \( z_2 = \textit{Proj}(\boldsymbol{h}_2) \), where \( z_1, z_2 \in \mathbb{R}^{D_p} \), and \( \mathbb{R}^{D_p} \) is the \( D_p \)-dimensional real-valued vector. 
\end{itemize}

\begin{figure*}
    \centering
    \includegraphics[width=0.88\linewidth]{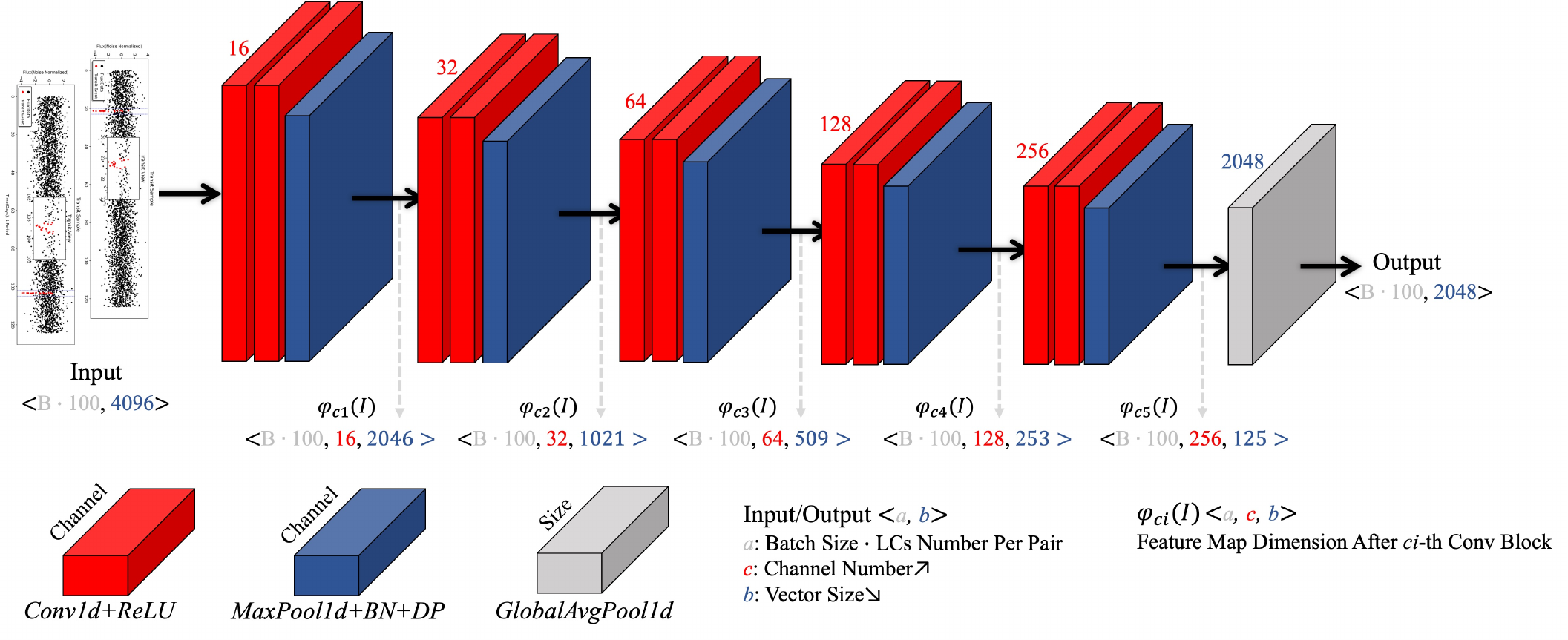}
    \caption{Encoder network architecture of DELOS. 
    The red cubes represent 1D \textit{Convolution layers} (\textit{Conv1d}) with \textit{ReLU} \textit{activation}, the blue cubes represent 1D \textit{Max Pooling} \textit{layers} (\textit{MaxPool1d}) with \textit{BatchNorm layer} (\textit{BN}) and \textit{Dropout} \textit{layer} (\textit{DP}), and the gray cubes represent 1D \textit{Global Average Pooling} \textit{layers} (\textit{GlobalAvgPool1d}). 
    The numbers on the slanted edge of each cube indicate information related to the output.
    Specifically, the Encoder network is composed of five convolutional blocks.
    We show the dimensions of feature map \(\varphi_{ci}(I)\)\(<\textit{a}\), \(c\), \(\textit{b}>\)obtained as input data \(I\) passes through the \(ci\)-th convolutional block, along with the input and output dimensions\(<\textit{a}\), \(\textit{b}>\), where \(\textit{a}\) refers to the batch size \(\cdot\) the number of light curves in each pair, \(\textit{b}\) refers to the number of data points in a light curve, and \(c\) refers to the number of channels in the feature map.}
    \label{fig:6figure}
\end{figure*}

Together, these components define how each original transit or noise sample is transformed into augmented views and then mapped into the representation space used by the supervised contrastive loss. 
The learned organization of this space is shown in Figure~\ref{fig:t-SNE}, where samples with the same label form compact structures, and transit and no-transit samples become separated.

\section{DELOS implementation}
\label{app:delos_implementation}
This appendix describes the implementation settings and network structures of the DELOS framework introduced in Section~\ref{sec:2.3}. 
The main text presents the overall training organization and end-to-end decision process, while the following paragraphs provide the corresponding implementation environment, batch configuration, classifier, projection network, and encoder architecture.

To implement this, we used the \href{https://pytorch.org/}{\texttt{PyTorch}} programming framework to develop DELOS and trained it on an NVIDIA 4090 graphics card.
The overall framework of DELOS is shown in Figure~\ref{fig:5figure}. 
Each batch input to DELOS consists of 32 pairs that form the basic unit for gradient computation and parameter updates. Meanwhile, the temperature parameter \( \tau \) of the loss is set to 0.07.
This framework is divided into two stages: the representation learning stage and the transit identification stage. The two stages update the parameters simultaneously during training but are driven by different losses.
In the former stage, we use the supervised contrastive loss of Equation~\ref{eq:scl} to optimize the encoder and projection networks. 
In the latter stage, we freeze the parameters of the encoder and projection networks and optimize only the classifier using the cross-entropy loss.
The classifier is composed of three fully connected layers, as shown in the following equation:
\begin{equation}
{\hat{y} = \sigma\left({W}^{(3)} \mathcal{R}\left({W}^{(2)} \mathcal{R}\left({W}^{(1)} \boldsymbol{h}\right)\right)\right)}
\label{eq:4}
\end{equation}
Here, \(\sigma(\cdot)\) is the softmax function, \(W^{*}\) is the weight matrix of each layer, and \(\mathcal{R}(\cdot)\) represents the ReLU activation. 
\(\hat{y}\) is a transit-likeness score in \((0,1)\), where higher values closer to 1 indicate stronger transit-like signatures in the folded light curve rather than a calibrated probability.

The DELOS projection network is also a fully connected, two-layer neural network, as shown in the following equation:
\begin{equation}
\boldsymbol{z} = W^{(2)} \mathcal{R}\left(W^{(1)} \boldsymbol{h}\right)
\label{eq:projection}
\end{equation}
Here, \(\boldsymbol{h}\) is the representation output by the Encoder, used for calculating the classification loss, and \(\boldsymbol{z}\) is the decoupled representation obtained through the Projection network, used for calculating the contrastive loss.

We customized the encoder network, which consists of five convolutional blocks, where the number of channels empirically increases while the vector size decreases as the data is processed sequentially.
The specific encoder architecture is continuously tuned to obtain the optimal encoder network, as illustrated in Figure~\ref{fig:6figure}, which can effectively capture both global and local multiscale representations of the light curves.
BatchNorm layer and Dropout layer were added after each MaxPooling layer to improve model convergence and mitigate overfitting. 
Furthermore, GlobalAveragePooling layer was applied at the end of the model to standardize the output to 2048 dimensions, ensuring robust generalization across varying input sizes.

\section{Binning strategy}
\label{app:optimal_binning}
An appropriate binning strategy is particularly critical for DELOS because it can suppress noise while preserving ITL, low-SNR transit signals potentially indicative of habitable exoplanets in \textit{Kepler} data. 
Specifically, when the number of bins is relatively small, weak transit signals may be smoothed out by background noise, resulting in a reduced transit SNR; 
in contrast, a relatively large number of bins improves temporal resolution but also leads to higher noise levels, a greater probability of missed detections, and increased computational overhead.

In the BLS algorithm paper, \cite{bls} stated that using a bin width smaller than the expected transit duration does not affect the detection performance.
However, if light curves are dominated by correlated noise \citep{bls_imp_4} or contain particularly weak signals, such strategies often lead to the signal being smoothed out, especially for Earth-like transit signals. 
Consequently, several subsequent studies \citep{bin_bls_imp_2, TLS, bin_bls_imp_3} convergently proposed a folded but binless BLS-like search strategy to preserve the full structure of such transit signals, although these approaches may naturally incur substantially higher computational overhead.
With the advancement of neural networks in the 2010s, a growing body of studies by \cite{astronet, wasp, astronet-k2, tess_1, ngts} has introduced AI-based approaches to identify transit signals.
These methods typically do not impose strict constraints on bin width, and \cite{tess_2} have even shown that using fewer bins can improve detection performance in the \textit{TESS} mission.
The use of relaxed bin width settings is reasonable because these approaches construct network inputs based on prior transit parameters derived from BLS or TLS, such as global, local, or half-phase views \citep{tess_4}.
Even if binning causes signal distortion in the global view, the local or half-phase view can still retain critical transit features, thereby mitigating the impact.

\begin{figure}
    \centering
    \includegraphics[width=0.9\linewidth]{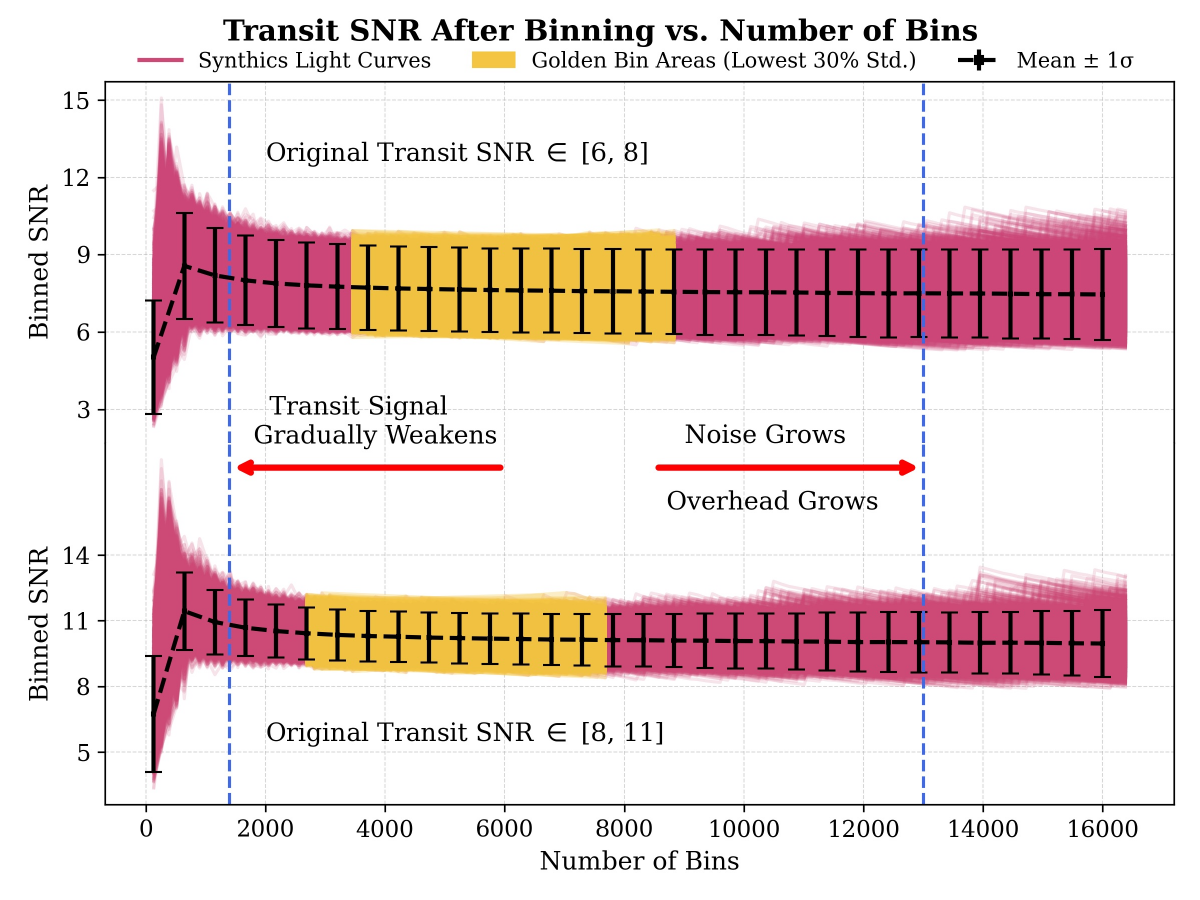}
    \caption{Variation of binned transit SNR with increasing binning number.
    The upper panel shows results for 4,000 synthetic light curves with original transit SNR $\in$ [6, 8], while the lower panel corresponds to 6,000 synthetic light curves with original transit SNR $\in$ [8, 11].
    The red curves represent the binned transit SNR variations for each individual light curve. 
    The yellow-shaded region marks the optimal binning region, also called the golden bin area, defined by the lowest 30\% $\sigma$ of the SNR.
    Black error bars indicate the $\pm1\sigma$ at each binning number.
    The results show that for ITL, low-SNR transits, the transit SNR becomes stable when the binning number exceeds a threshold of about 2000.}
    \label{fig:bin_1_figure}
\end{figure}

\begin{figure*}
    \centering
    \includegraphics[width=0.88\linewidth]{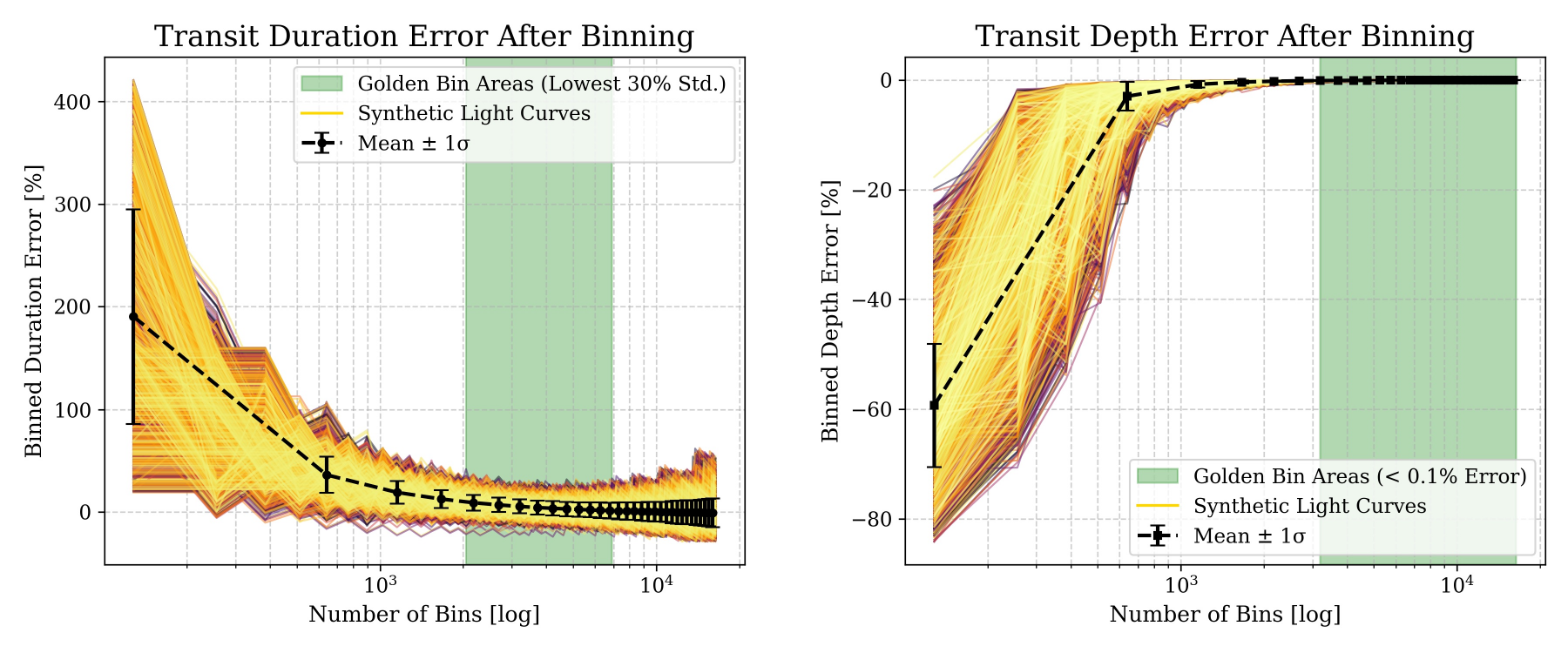}
    \caption{Dependence of transit-duration and transit-depth preservation on the adopted phase-binning number.
    The relative error is defined as $(\text{binned value} - \text{original value}) / \text{original value} \times 100\%$, with values closer to zero indicating better preservation of the original transit morphology after binning. The left panel shows the relative error in transit duration, while the right panel shows the relative error in transit depth, measured for 10,000 simulated light curves.
    The yellow curves represent individual simulated light curves, and the black error bars indicate the $\pm 1 \sigma$ scatter at each binning number.
    The green shaded regions mark the optimal binning intervals: for transit duration, the region corresponding to the lowest 30\% of the scatter in relative error; for transit depth, the region where the relative error remains below 0.1\%.
    These results show that both transit-duration and transit-depth information are well preserved once the binning number is sufficiently large, supporting the choice of 4096 bins adopted in DELOS.}
    \label{fig:bin_2_figure}
\end{figure*}

However, this reliance on traditional algorithmic priors imposes a critical limitation: the overall detection capability of the neural network becomes bounded by the performance ceiling of BLS or TLS. 
In both the periodic detection algorithms mentioned above and the end-to-end models, the low-SNR nature of transit signals remains one of the most critical and challenging issues.
Fortunately, binning can improve the initial SNR of transit signals to some extent, thereby increasing the detection sensitivity of algorithms to low-SNR events.
Accordingly, we systematically investigate the optimal number of bins from two perspectives.
\begin{itemize}
\setlength{\itemsep}{2pt}
\setlength{\topsep}{0pt}
    \item Signal preservation: The duration and depth of the binned transit should remain consistent with those of the original signal, introducing no significant distortion or bias.
    \item Transit SNR trade-off: The binning should suppress noise effectively while ensuring that the resulting SNR is not less than that of the original signal and ideally achieves a measurable improvement.
\end{itemize}

Specifically, we adopt the data simulation pipeline described in Section~\ref{sec:3.2} to generate a synthetic dataset consisting of 10,000 light curves, each mimicking a four year Kepler-like observational baseline.
The transit signals embedded in these light curves have uniformly distributed SNRs in the range [6, 11], with corresponding orbital periods in the range [100, 150] days.
The noise data for all synthetic light curves are derived from real \textit{Kepler} data.
Each light curve is first folded in its true orbital period and then binned into a range of numbers, from 128 to 16,384 in steps of 128.

We recorded changes in transit SNR, duration, and depth before and after binning the folded light curves to evaluate the impact of binning on signal characteristics.
To more intuitively demonstrate the sensitivity of Low SNR transits to binning size, we divided the 10,000 simulated light curves into two subsets based on their SNR values: an Extremely Low SNR group (ELS) with SNR $\in$ [6, 8] and a Low SNR Group (LSG) with SNR $\in$ [8, 11].
Figure~\ref{fig:bin_1_figure} illustrates how the binned transit SNR evolves with the binning numbers for both the ELS and LSG groups.
In this Figure, when the number of bins is below 2000, the transit signal is largely smoothed out by background noise, resulting in high variance of transit SNR, sometimes even lower SNR than the original unbinned transit.

To balance signal stability and computational efficiency, we highlight in Figure~\ref{fig:bin_1_figure} the binning intervals corresponding to the lowest 30\% $\sigma$ of the SNR, which we define as the optimal binning region.
In particular, the ELS and LSG groups exhibit different behaviors in this regard.
Since LSG signals are inherently stronger than those in ELS, their SNR $\sigma$ is generally smaller across binning scales, and their optimal binning region tends to lie at lower binning numbers. 
Specifically, the optimal binning range for ELS is [3456, 8832], while that for LSG is [2688, 7680].

In addition, Figure~\ref{fig:bin_2_figure} presents the change in error for binned transit duration and depth relative to their original values across different binning numbers for the entire set of 10,000 light curves.
The left panel of Figure~\ref{fig:bin_2_figure} shows that the binned duration error exhibits noticeable $\sigma$ depending on the number of bins, with a trend similar to the physical mechanism described in Figure~\ref{fig:bin_1_figure}.
Based on the interval with the 30\% smallest $\sigma$, we calculate its optimal binning region to be approximately [2048, 6912].
In the right panel of Figure~\ref{fig:bin_2_figure}, the transit depth error rapidly converges to near zero once the binning number exceeds around 2000.
Consequently, we define the optimal binning region for depth error as the binning range in which the error remains below 0.1\%, corresponding to [3200,10000).

By overlaying the four optimal binning regions discussed above, we identify the range [3456, 6912] as the most stable interval for the enhancement of transit SNR after binning, while preserving the morphology of the transit.
The 4096 binning number adopted by DELOS falls precisely within this interval, further validating the effectiveness of binning strategies in detecting weak Earth-like transits.
Moreover, as the target search period increases, the optimal binning number tends to shift toward higher values.
This trend provides meaningful guidance for future applications of DELOS, GPFC, or similar methods to efficiently identify low-SNR transits across arbitrary periods.

As noted in \cite{undrending_2}, encoder networks such as the one shown in Figure~\ref{fig:6figure} inherently perform a binning-like operation through pooling-layer downsampling when extracting temporal representations from light curves.
Based on this insight, we argue that it is preferable to select binning numbers toward the upper end of the optimal binning region to ensure that the encoder retains more in-transit data points during the downsampling process, thereby improving the effectiveness of feature extraction.
Importantly, the optimal encoder architecture should follow several principles: progressively increasing the number of channels layer by layer, maintaining sufficiently dense sampling along the one-dimensional feature vector, and minimizing the span between the input and output vectors.
These designs are particularly crucial for extracting sparse transit features, and they can significantly enhance the representational capacity of the model in detecting low-SNR transit signals with longer orbital periods.

\section{Runtime and convergence}
\label{app:runtime_convergence}
This appendix summarizes the runtime information for synthetic data generation and DELOS training, along with the recorded training convergence curves referenced in Section~\ref{sec:3.3}.

Our parallel program was executed on \textit{the Ubuntu} operating system, equipped with \textit{an AMD Ryzen 9 7950X 16 core processor}, requiring approximately 6 to 7 hours to generate the 200,000 pairs.

The training loss and accuracy were recorded throughout the optimization process to monitor convergence.

\begin{figure}
    \centering
    \includegraphics[width=0.8\linewidth]{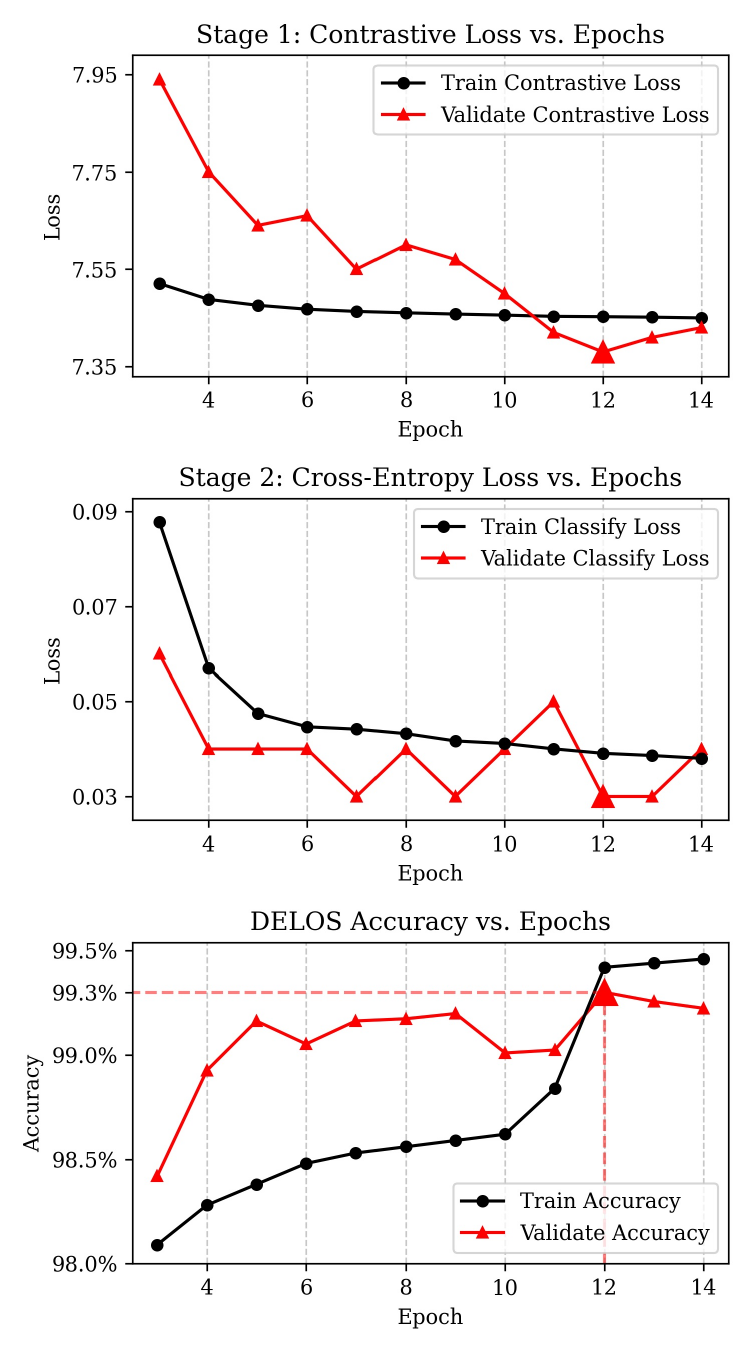}
    \caption{Variations in DELOS loss and accuracy versus epochs during training. 
    The bold red triangle highlights the epochs where DELOS achieved best performance.}
    \label{fig:9figure}
\end{figure}

Trends in loss and accuracy throughout the training process were recorded, as shown in Figure~\ref{fig:9figure}.
DELOS was implemented on \textit{the Ubuntu} operating system powered by an NVIDIA 4090 graphics card, completing 14 epochs in approximately 6 hours.

\section{inference profiling}
\label{app:inference_profiling}
This appendix summarizes the inference procedure of DELOS and the GPU profiling analysis used to determine the \textit{BatchSize} adopted in Section~\ref{sec:4.3}.
The process by which DELOS produces the score periodogram is detailed in the Algorithm~\ref{alg:DELOS}.

\begin{algorithm}[H]
\caption{DELOS for Detecting Transits}
\label{alg:DELOS}
\small
\begin{algorithmic}[1]
\REQUIRE lc, periods, device, batchSize
\ENSURE results (list of predicted results)

\STATE \textbf{Preliminary Step: Folded and Normalized Data}
\STATE foldLc $\gets$ GpuFold(lc, periods, device)
\STATE data $\gets$ BinAndNoiseNormalize(foldLc)

\vspace{0.5em}
\STATE \textbf{Step 1: Load Data onto GPU}
\STATE gpuData $\gets$ LoadToGpu(data)

\vspace{0.5em}
\STATE \textbf{Step 2-5: Iterate Over Batches}
\STATE $i$ $\gets$ $0$
\WHILE{$i < 58,400 / batchSize$}
    \STATE \textbf{Step 2: Slice+Reshape Data}
    \STATE batchData $\gets$ Deal(gpuData, i, batchSize)
     
    \vspace{0.3em}
    \STATE \textbf{Step 3: Feature Extract}
    \STATE feature $\gets$ ExtractFeatures(batchData, Extractor)
    
    \vspace{0.3em}
    \STATE \textbf{Step 4: Transit Classify}
    \STATE batchPredictions $\gets$ Classify(feature, classifier)
    
    \vspace{0.3em}
    \STATE \textbf{Step 5: Collect Results}
    \STATE results $\gets$ AppendResults(batchPredictions, results)
    
    \vspace{0.3em}
    \STATE $i$ $\gets$ $i + 1$
\ENDWHILE

\vspace{0.5em}
\RETURN results
\end{algorithmic}
\end{algorithm}

The score-periodogram generation consists of two main stages: GPU-Fold in the preliminary step and contrastive-learning inference in Steps~1--5.
During the latter stage, the computational speed is strongly influenced by the \textit{BatchSize} hyperparameter. 
To investigate this, we conducted experiments that analyzed DELOS speed under varying \textit{BatchSize} settings, focusing on minimizing the time cost of key steps and maximizing two GPU metrics from \cite{tensorflow2015-whitepaper}: GPU utilization ($U_{\text{GPU}}$) and estimated Streaming Multiprocessor (SM) efficiency ($\eta_{\text{SM}}$).
We monitored these metrics using \href{https://www.tensorflow.org/tensorboard}{\texttt{TensorBoard}} software, and the calculation formulas for $U_{\text{GPU}}$, $\eta_{\text{SM}}$ are presented below:
\begin{equation}
U_{\text{GPU}} = \frac{T_{\text{busy}}}{T_{\text{total}}}
\label{eq:U_gpu}
\end{equation}
Here, \(T_{\text{busy}}\) represents the GPU busy time, and \(T_{\text{total}}\) refers to the total execution time. Consequently, it measures the activity level of the GPU during task execution, reflecting its overall workload.
\begin{equation}
\eta_{\text{SM}} = \frac{1}{T_{\text{total}}} \cdot\sum_{k=1}^{K} \left(\min\left(\frac{N_{\text{blocks}, k}}{N_{\text{SM}}}, 1.0\right) \cdot T_{\text{exec}, k} \right)
\label{eq:E_SM}
\end{equation}
Here, \(N_{\text{blocks}, k}\) represents the number of thread blocks allocated to the \(k\)-th kernel, \(N_{\text{SM}}\) denotes the number of SMs in the GPU, and \(T_{\text{exec}, k}\) is the execution time of the \(k\)-th kernel.
Consequently, it quantifies the utilization of computational resources within each SM, indicating the effectiveness of the thread-block distribution.

Generally, larger values of \(U_{\text{GPU}}\) and \(\eta_{\text{SM}}\) result in faster speeds. However, excessively large values may introduce new issues, increasing overall runtime.
For example, data transfer between the CPU and GPU can become a bottleneck; improper thread block size or shared memory configuration may lead to register or shared memory becoming performance bottlenecks; and task scheduling and context switching can add additional overhead. 
Therefore, identifying the \textit{sweet spot} for DELOS in the light curve evaluation process is crucial, i.e., exploring the optimal \textit{BatchSize} value.
We conducted experiments on an \textit{Ubuntu} operating system with two NVIDIA 4090 graphics cards and \textit{AMD Ryzen 9 7950X 16-core processor}, ensuring that all other processes were terminated. 
Moreover, all experiments were repeated 50 times with the same \textit{BatchSize} configuration.

Finally, we found that when the \textit{BatchSize} is set to 128, the time cost during the contrastive learning inference stage broke the one-second barrier, achieving 0.97 seconds. At the same time, \(U_{\text{GPU}}\) and \(\eta_{\text{SM}}\) were approximately 91 percent.
However, beyond this value, the overall runtime increased, as shown in Figure~\ref{fig:15figrue}. 
Therefore, we identify \textit{BatchSize} = 128 as the \textit{sweet spot} in the DELOS detection system and maintain it in subsequent experiments.

\end{appendix}

\end{document}